\documentclass[letterpaper,10pt,final,balancelastpage,floats,aps,twocolumn]{revtex4}
\usepackage{amsmath}
\usepackage{subfigure}
\usepackage{graphicx}% Include figure files
\usepackage{dcolumn}% Align table columns on decimal point
\usepackage{bm}% bold math
\usepackage{algorithm}
\usepackage{algorithmic}

\begin{document}
\newcommand{\italics}{\textit}
\newcommand{\etal}{{\textit{et al}. }}

\title{Orbital Optimization in the Density Matrix Renormalization
  Group, with applications to polyenes and $\beta$-carotene}% Force line breaks with \\

\author{Debashree Ghosh}
\affiliation
{
Department of Chemistry and Chemical Biology, Cornell University,
Ithaca, New York 14853-1301, USA
}
\author{Johannes Hachmann}
\affiliation
{
Department of Chemistry and Chemical Biology, Cornell University,
Ithaca, New York 14853-1301, USA
}
\author{Takeshi Yanai}
\affiliation
{
  Department of Theoretical and Computational Molecular Science,
  Institute for Molecular Science, Okazaki, Aichi 444-8585,
  Japan
}
\author{Garnet Kin-Lic Chan}
\affiliation
{
Department of Chemistry and Chemical Biology, Cornell University,
Ithaca, New York 14853-1301, USA
}
 \email{gc238@cornell.edu}

\date{\today}% It is always \today, today,

\begin{abstract} 
In previous work we have shown that the  Density Matrix
Renormalization Group (DMRG) enables  near-exact calculations in 
active spaces much larger than are possible with traditional Complete
Active Space algorithms. Here, we implement orbital
optimisation with the Density Matrix Renormalization Group to further allow
the self-consistent improvement of the active orbitals, as is 
done in the  Complete Active Space Self-Consistent Field (CASSCF) method. 
We use  our  resulting DMRG-CASSCF method to study the low-lying excited states
of the \textit{all-trans} polyenes up to $\text{C}_{24}\text{H}_{26}$ as well as $\beta$-carotene,  correlating with near-exact
accuracy the optimised  complete $\pi$-valence space with up to 24
active electrons and orbitals, and analyse
our results in the light of the recent discovery from Resonance Raman
experiments of new  optically dark states in the spectrum.
\end{abstract}

\maketitle
\section{Introduction}

\label{sec:intro}
The Density Matrix Renormalization Group (DMRG) is an electronic
structure method that has recently been applied to  \textit{ab-initio} quantum chemistry. The method originated in the condensed matter community
with the pioneering  work of White
\cite{White1992,White1993}. Although the earliest quantum
chemistry implementations are only a few years old, the DMRG has already
enabled the solution of many problems  that would be intractable with any
other method \cite{White1999,Mitrushenkov2001,Chan2002,Legeza2003dyn,Moritz2007}. For example, we have shown that the DMRG can
obtain near-exact solutions to multireference problems with active
spaces much larger than are possible with traditional active space techniques. 
Such problems have ranged from  molecular potential energy
curves \cite{Chan2003,Chan2004b}, to the ground and excited states of large conjugated polymers \cite{Hachmann2006,Dorando2007,Hachmann2007}, to
metal-insulator  transitions in hydrogen chains
\cite{Hachmann2006}. In each of these cases, we obtained  DMRG energies
within $0.001$-$0.1$m$E_h$ of the (estimated) exact Full Configuration
Interaction (FCI) energies in the active space, but for active spaces that,
in some problems, have been as large as 100 active electrons in 100  orbitals \cite{Hachmann2006}. 
The development of the DMRG in quantum chemistry has
proceeded through the efforts of several groups, and we mention here
the  work of White \etal  \cite{White1999,Daul2000,Rissler2006},  Mitrushenkov \etal  \cite{Mitrushenkov2001,Mitrushenkov2003,Mitrushenkov2003nort}, our
contributions \cite{Chan2002,Chan2003,Chan2004,Chan2004b,Chan2005,Hachmann2006,Dorando2007,Hachmann2007}, the work of Legeza,
Hess \etal \cite{Legeza2003dyn,Legeza2003qie,Legeza2003lif,Legeza2004}, the work
of Reiher \etal \cite{Moritz2005orb,Moritz2005rel,Moritz2006,Moritz2007}, and most
recently the work of Zgid and Nooijen \cite{Zgid2008}. Also related, but too numerous to cite
in full here, are earlier developments  of the method for  semi-empirical Hamiltonians; some representative contributions are those
in Refs. \cite{Ramasesha1997,Yaron1998,Shuai1998,Fano1998,Bendazzoli1999,Raghu2002a,Raghu2002b}.

At the heart of the DMRG is a  wavefunction ansatz  and the 
DMRG ``algorithm'' is simply an efficient variational optimisation
procedure for this ansatz. Unlike most wavefunctions  in quantum chemistry, the
DMRG wavefunction is not parametrised by excitations from an underlying reference
state. Rather, it is  built directly from local variational 
objects (which we shall later call \textit{site functions}) which are associated with the active orbitals in the
system, and which describe how the orbitals are correlated with each
other. Each site function is characterised by
a rank $M$ that measures the number of variational parameters,  and
as this rank increases the ansatz becomes exact. 
For an incomplete rank $M$, correlations between orbitals that
are widely separated in the ansatz are truncated. Thus the DMRG is a
naturally local theory,  but, since the  ansatz is not constructed
from a reference, it is a local \textit{multireference} theory. This
may be seen as the basic reason why the DMRG can describe very  large multireference
problems so easily. We should note  that the structure of the
DMRG wavefunction means that it is  a local theory only in 
the number of correlating orbitals along \textit{one} of the physical
dimensions of the problem. However, generalisations of the ansatz to a
local theory along \textit{all} physical dimensions are now known, and
are under active development \cite{Verstraete2004pbc,Verstraete2004peps, Perez-Garcia2007,Schuch2007,Murg2007,Vidal2006}.

In most applications of the DMRG to  quantum chemistry so far,  the active
space of interest has been easy to identify, i.e. there is a good
core-valence and valence-Rydberg separation, either for energetic or
symmetry reasons, allowing the DMRG to be used with such an active
space as a direct substitute for Complete Active Space Configuration Interaction (CASCI). 
 In general, however, we cannot always identify the active orbitals in a
 simple way, and thus there is a need for an \textit{orbital
optimised} DMRG, where the active space is determined self-consistently
by energy minimisation, in much the same way as in the Complete Active Space Self-Consistent Field
(CASSCF) method \cite{Roos1986,Roos1987}. The purpose of the current work is to describe how
this may be done. The resulting orbital optimised
DMRG we shall refer to as the \textbf{DMRG-CASSCF} method.

While the general idea of orbital optimisation is straightforward, in
practice an efficient implementation must be tailored to the
underlying many-body wavefunction ansatz. In 
Sec. \ref{sec:theory} we describe such an algorithm for the DMRG
wavefunction. We start with an overview of orbital optimisation in
Sec. \ref{sec:overview} that recalls how the procedure may naturally be
divided into two  tasks, the evaluation of the one- and
two-particle density matrices, and the orbital
rotation and integral transformation steps. In
Sec. \ref{sec:structure} we present an efficient method to evaluate the
 one- and two-particle density matrices in the DMRG. 
Our current implementation benefits from the observation of Zgid and
Nooijen that the one-site DMRG algorithm is more suitable than the two-site
DMRG algorithm  for this purpose. 
To facilitate the large-scale calculations for our
applications to long polyenes and $\beta$-carotene in this work, we
have fully parallelised not only the evaluation of the reduced density
matrices in the DMRG, but
also  the orbital rotation and integral transformation steps. These
implementation aspects are discussed in Sec. \ref{sec:orbital}. Finally, the complete DMRG-CASSCF
macroiteration is summarised in Sec. \ref{sec:complete}.

In Sec. \ref{sec:apps} we apply the
DMRG-CASSCF method to the problem of the low-lying excitations in
polyenes and $\beta$-carotene. The conjugated $\pi$-system
in the polyenes and  substituted species such as
$\beta$-carotene gives rise to an unusual excitation spectrum, with
``dark'' electronic states lying beneath the optically allowed HOMO-LUMO
transition. The electronic structure of these low-lying states lies at
the heart of  energy transport in systems ranging from conjugated  organic
semiconductors to the biological centres of light-harvesting and
vision.  While the relevant active space on these systems clearly 
consists of  the conjugated $\pi$-valence orbitals, to the best of our knowledge  previous
calculations on these systems have not correlated  complete
$\pi$-valence spaces with more than 5 double bonds (corresponding to
a (10,10) complete active space \cite{hirao,kurashige}). In the
current study we use our  DMRG-CASSCF method to perform
calculations correlating the complete $\pi$-valence space in polyenes
up to $\text{C}_{24}\text{H}_{26}$ (with 12 conjugated bonds)
and $\beta$-carotene (with 11 conjugated bonds), and analyse our
results in relation to recent Resonance Raman measurements, which
have detected  previously unidentified  ``dark'' states in the low-lying spectrum.

\section{Theory}

\label{sec:theory}
\subsection{Overview of orbital optimisation}

\label{sec:overview}
We begin with some general remarks on orbital optimisation in
\textit{ab-initio} quantum chemistry.
Starting from the electronic Hamiltonian, specified by the one- and
two-electron integral matrix elements $t_{ij}$ and $v_{ijkl}$
\begin{equation}
H = \sum_{ij}t_{ij} a^\dag_i a_j
  + \sum_{ijkl} v_{ijkl} a^\dag_i a^\dag_j a_k a_l \label{eq:qc_H}
\end{equation}
an \textit{ab-initio} quantum chemical method  provides a wavefunction
$\Psi$ that approximates a target  eigenstate of $H$. From
$\Psi$  we  define the  one- and two-particle density matrix elements
$\gamma_{ij}, \gamma_{ijkl}$ 
\begin{align}
\gamma_{ij} &= \langle \Psi | a^\dag_i a_j| \Psi \rangle \\
\gamma_{ijkl} &= \langle \Psi | a^\dag_i a^\dag_j a_k a_l| \Psi \rangle 
\end{align}
and the energy  expectation value $\langle \Psi|H|\Psi\rangle$ can be
written as
\begin{equation}
E =  \sum_{ij} t_{ij} \gamma_{ij}
  + \sum_{ijkl} v_{ijkl} \gamma_{ijkl} \label{eq:qc_Henergy}
\end{equation}

Orbital rotation corresponds to a unitary transformation of
the wavefunction effected by an operator $e^A$, where $A$ has
the single-particle operator form
\begin{equation}
A = \sum_{ij} A_{ij} a^\dag_i a_j \label{eq:exp_rot}
\end{equation}
and $A_{ij} = -A_{ji}^*$. 
After orbital
rotation, the transformed wavefunction $\bar{\Psi}$ and energy
$\bar{E}$ are
\begin{align}
\bar{\Psi}& = e^A \Psi \nonumber \\
\bar{E} &= \langle \Psi e^{-A} | H | e^A \Psi \rangle \label{eq:rotated_energy}
\end{align}
But one can also consider the unitary operator to act on the Hamiltonian rather
than the wavefunction, and from this equivalent point of view, we  have a transformed
$\bar{H}$  and energy expression
\begin{align}
\bar{H}& = e^{-A} H e^A\nonumber \\
\bar{E}& = \langle \Psi|  \bar{H} |\Psi \rangle
\end{align}
The transformed Hamiltonian $\bar{H}$ has the same form as the original Hamiltonian (\ref{eq:qc_H}) but with modified
integrals  $\bar{t}_{ij}$ and $\bar{v}_{ijkl}$ that reflect the
rotated orbitals
 \begin{align}
 \bar{t}_{ij} &= \sum_{i'j'} U_{ii'}^* U_{jj'} {t}_{i'j'} \nonumber \\
 \bar{v}_{ijkl} &= \sum_{i'j'k'l'} U_{ii'}^* U_{jj'}^* U_{kk'} U_{ll'} {v}_{i'j'k'l'} \label{eq:mo_integral_transform}
 \end{align}
where $U$ is the coefficient matrix $e^A$.
 Thus we can rewrite the energy after orbital rotation in terms of the original one- and two-particle density
matrices and the modified integrals 
\begin{align}
% \tilde{E} = \tilde{E}^\mathrm{closed}
%           + \sum_{ij} \tilde{t}^\mathrm{act}_{ij} \gamma_{ij}
%           + \sum_{ijkl} \tilde{v}_{ijkl} \gamma_{ijkl} \label{eq:rotated_energy_terms}
\bar{E} = \sum_{ij} \bar{t}_{ij} \gamma_{ij}
          + \sum_{ijkl} \bar{v}_{ijkl} \gamma_{ijkl} \label{eq:rotated_energy_terms}
\end{align}

We include this elementary discussion because it leads directly to the
following familar procedure to optimise the orbitals in an \textit{ab-initio} wavefunction:
\begin{enumerate}
\item From the \textit{ab-initio} method  obtain $\Psi$ corresponding
  to the given $H$ and form the density matrices  $\gamma_{ij}, \gamma_{ijkl}$.
\item Determine an orbital rotation step $e^A$, and form the new Hamiltonian
  $\bar{H} =e^{-A} H e^A$ from the transformed integrals.
\item Goto 1. and loop until convergence in $\Psi$.
\end{enumerate}
Note that in the above, the orbital degrees of freedom and the
other  ansatz degrees of freedom in $\Psi$ are alternately 
optimised in steps (1), (2). While more sophisticated approaches
which couple orbital rotations with changes in the other ansatz degrees of
freedom can be envisaged (as are employed in multi-configurational
self-consistent field methods \cite{knowles,yeager1982nra}), we shall adopt  the above simple strategy  to optimise the orbitals in the DMRG wavefunction.
The conceptual task is then twofold. Firstly,  how do we  calculate the one- and two-particle
density matrices in the DMRG? And secondly, what method should we use to  select our orbital
rotation steps and to construct the transformed Hamiltonian?

\subsection{Evaluation of the one- and two-particle
  density matrices in the DMRG}

\label{sec:structure}
% \input{header.tex}

% \begin{document}

\newcommand{\boldpsi}{\boldsymbol{\psi}}

While the algorithm to calculate the one- and two-particle density
matrices could, in principle, be described entirely in the traditional
Renormalization Group language of the DMRG, we believe that it is
beneficial to understand the method in  a more modern
language which focuses on the  structure of 
the DMRG wavefunction. Thus we begin with a brief review of the
general properties of the DMRG wavefunction before proceeding to the
method of reduced density matrix evaluation. For an expanded introduction to the wavefunction perspective in
DMRG, we refer the reader to our  
introductory article Ref. \cite{Chan-bookchapter}
as well as  other  recent reviews in the field \cite{SCHOLLWOCK:2005:_dmrg}.

% While we shall not present a complete discussion of
% the DMRG wavefunction here, we refer the reader to our recent
% introductory article 

\subsubsection{ The  DMRG wavefunction } 
% Before describing the algorithm to determine the one- and two-particle
% density matrices in the DMRG, we  will need to recapitulate some
% basic features of the DMRG ansatz. (We do not attempt a complete discussion of this interpretation of the DMRG
% here but refer the reader instead to our recent introductory article 

The DMRG algorithm corresponds to 
 a variational minimisation of the energy   within the space of a wavefunction ansatz. To specify this ansatz
 we  first define  an ordering of the orbitals thereby mapping them onto sites on a one-dimensional
lattice. Then, the ``one-site''  DMRG ansatz is given by
\begin{align}
|\Psi_\text{DMRG}\rangle = \mathop{\sum_{n_1 n_2 n_3
    \ldots n_k}}_{i_1i_2i_3 \ldots i_{k-1}} \psi^{n_1}_{i_1}
\psi^{n_2}_{i_1 i_2} \psi^{n_3}_{i_2 i_3} \ldots \psi^{n_k}_{i_{k-1}} |n_1
n_2 n_3 \ldots n_k\rangle \label{eq:dmrg_ansatz}
\end{align} 
where $|n_1 \ldots n_k\rangle$ denotes a Slater determinant in occupation
number form, i.e.   $n_i$ is the occupation of orbital $i$, and the total number of
orbitals is $k$. The $\psi$ ``site functions''   are 3-index quantities
and are the variational parameters of the wavefunction. 
The dimension of each $n_1 \ldots n_k$ index is 4, corresponding to the 4 occupancies for each orbital
$|-\rangle, |\phi^\alpha\rangle, |\phi^\beta\rangle, |\phi^\alpha\phi^\beta\rangle$,
while the dimension of each auxiliary index $i_1 \ldots
i_{k-1}$ is some specified size $M$, thus making each site function a tensor of dimension $4
\times M \times M$, except for the first and last, which only have two
indices and are of dimension $4 \times M$. As $M$
increases, the wavefunction ansatz becomes increasingly exact. 
If we  interpret a site function with indices $n_p, i_{p-1}, i_p$ as  a \textit{matrix array} $\boldpsi^{n_p}$ where
$i_{p-1}, i_p$ are the matrix indices and $n_p$ is the third, array,
index,  then the ansatz is written  compactly as a matrix product state
\begin{align}
 |\Psi_\text{DMRG}\rangle = \mathop{\sum_{n_1 n_2 n_3
     \ldots n_k}} \boldpsi^{n_1}
 \boldpsi^{n_2} \boldpsi^{n_3} \ldots \boldpsi^{n_k} |n_1
 n_2 n_3 \ldots n_k\rangle
\end{align} 
Because of this  matrix product structure, the DMRG ansatz is also known as
the matrix product state (MPS) \cite{Fannes1992,Ostlund1995,Rommer1997}.

% will find it useful to use several notations for
% the site functions. For example, we can write  Alternatively, we can write it as a single compound-indexed  matrix  $\boldpsi^p$, where  
% the matrix indices are given by the grouping $i_{p-1}n_p, i_p$ or
% $i_{p-1}, n_p i_p$.  In the matrix array notation, 

%% e.g. for $\psi^{n_p}_{i_{p-1} i_p}$  index $n_p$ is an occupation
%% index for orbital $p$, and $i_{p-1}, i_p$ are called auxiliary indices
%% and describe the correlations between the different orbital
%% spaces. The dimension of the index $n_p$ is 4, corresponding to the 4
%% orbital states 
%% In the DMRG wavefunction with $M$ states, the dimensions of the
%% auxiliary indices is $M$, making the $\psi$ site functions 3-tensors of
%% dimension $4 \times M \times M$. 

%% Each site function (as a 3-tensor)
%% may be viewed as an array of matrices i.e. $\psi^{n_p}_{i_{p-1} i_p} =
%% [\psi^{n_p}]_{i_{p-1} i_p}$, and thus we can use the more compact
%% notation

\newcommand{\boldL}{\boldsymbol{L}}
\newcommand{\boldR}{\boldsymbol{R}}
\newcommand{\boldC}{\boldsymbol{C}}
\newcommand{\boldM}{\boldsymbol{M}}

\newcommand{\lpp}{l^\prime}
\newcommand{\npp}{n^\prime}
\newcommand{\rpp}{r^\prime}

Now the above form of the DMRG ansatz is  invariant
to transformations of the site functions  of the form ($\boldpsi^{n_p} \to
\boldpsi^{n_p} \mathbf{U}$, $\boldpsi^{n_{p+1}} \to \mathbf{U^\dag}
\boldpsi^{n_{p+1}}$) and thus it is useful to  define a \textit{canonical} form of the DMRG
wavefunction that eliminates this freedom. In practice, this canonical
representation is used  in all DMRG
calculations, and it is also the representation in which the link
between the DMRG wavefunction and the traditional Renormalization Group
language is most direct. In essence, the canonical form of the
wavefunction at a given site corresponds to the familiar expression for
the DMRG wavefunction where it is expanded in the product basis of the left and right
blocks separated by the site \cite{White1999,Chan-bookchapter}.

To obtain the canonical form, we
 choose a specific site, say $p$, around
which to canonicalise. Then the site $p$ canonical form is given as
\begin{align}
|\Psi \rangle& = \sum_{n_1 \ldots n_p \ldots n_k} {\boldL}^{n_1} \ldots {\boldL}^{n_{p-1}}
 {\boldC}^{n_p} {\boldR}^{n_{p+1}} \ldots {\boldR}^{n_k} |n_1 \ldots n_p
 \ldots n_k\rangle \label{eq:dmrg_mps}
%% \\
%% &=\mathop{\sum_{n_1 \ldots n_p \ldots n_k}}_{l_1
%%  \ldots l_{p-1}, r_p \ldots r_{k-1}} L^{n_1}_{l_1} \ldots L^{n_{p-1}}_{l_{p-2} l_{p-1}}
%%  C^{n_p}_{l_{p-1} r_p} R^{n_{p+1}}_{r_p r_{p+1}} \ldots R^{n_k}_{r_{k-1}} |n_1 \ldots n_p
%%  \ldots n_k\rangle . 
\end{align} 
We label   the site functions to the left of $p$  by 
$L$, and those to the right by $R$. The degeneracy (invariance to
transformation) of the original  ansatz (\ref{eq:dmrg_ansatz}) mentioned above is lifted by requiring the $L$ and $R$ site functions to be  orthogonal
projection matrices in the following sense
\begin{align}
\sum_{ln_q} L^{n_q}_{ll^\prime} L^{n_q}_{ll^{\prime\prime}} &=
\delta_{l^\prime l^{\prime\prime}}   \label{eq:lorth}\\
\sum_{rn_q} R^{n_q}_{r^\prime r} R^{n_q}_{r^{\prime\prime} r}& =
\delta_{r^\prime r^{\prime\prime}}  \label{eq:rorth} 
\end{align}
i.e. by grouping together the $ln_q$ indices to form the row index of a $4M
\times M$ matrix, each $L$ site function is orthogonal with respect to
its $M$ columns, while by grouping together the $rn_q$ indices to form
the column index of a $M \times 4M$ matrix,  each $R$ site function
is orthogonal with respect to its $M$ rows.
% \begin{align}
% \end{align}
% i.e. each $R$ site function is orthogonal with respect to the $M$ rows
% $r^\prime, r^{\prime\prime}$.

The link between the canonical form and the original RG
formulation appears when we combine the $L$ site functions $\boldL^{n_1} \ldots
\boldL^{n_{p-1}}$ with the basis states $|n_1 \ldots n_{p-1}\rangle$, and
the $R$ site functions $\boldR^{n_{p+1}} \ldots \boldR^{n_k}$ with the
basis states $|n_{p+1} \ldots n_{k}\rangle$, to define renormalised left
and right many body spaces $\{ l_{p-1} \}$, $\{ r_{p+1}\}$
\begin{align}
|l_{p-1} \rangle  &= \mathop{\sum_{n_1 \ldots n_{p-1}}}_{l_1 \ldots l_{p-2}}
           L^{n_1}_{l_1}   \ldots L^{n_{p-1}}_{l_{p-2}l_{p-1}} 
           |n_1  \ldots n_{p-1}\rangle\label{eq:basis_lq} \\
|r_{p+1} \rangle  &= \mathop{\sum_{n_{p+1} \ldots n_{k}}}_{r_{p+2} \ldots r_{k}}
           R^{n_{p+1}}_{r_{p+1} r_{p+2}}   \ldots R^{n_{k}}_{r_{k}}
           |n_{p+1}  \ldots n_{k}\rangle\label{eq:basis_rq}
\end{align}
Since the dimension of the  left basis in Eq. (\ref{eq:basis_lq}) is $M$ (i.e. the dimension of the
auxiliary index  $l_{p-1}$) and similarly for the right basis, the
site functions $\boldL^{n_1} \ldots
\boldL^{n_{p-1}}$ and $\boldR^{n_{p+1}} \ldots \boldR^{n_k}$ define a projective transformation or
\textit{renormalization} from the many-body spaces $\{ n_1 \} \otimes \ldots \otimes \{
n_{p-1}\}$ and $\{ n_{p+1}\} \otimes \ldots \otimes
\{ n_k \}$ to  the left and right spaces, $\{l_{p-1}\}, \{
r_{p+1} \}$, respectively. Then, in the renormalised representation,
$C^{n_p}_{l_{p-1} r_p}$ gives the coefficients of expansion of the
wavefunction $|\Psi\rangle$, i.e.
\begin{align}
|\Psi \rangle = \sum_{l_{p-1} n_p r_p} C^{n_p}_{l_{p-1} r_p} |l_{p-1} n_p r_p\rangle \label{eq:blockwf}
\end{align} 
This is just the RG expression for the one-site DMRG wavefunction,
in the product space of a renormalised left ``block'', a site $p$, and
a renormalised right ``block''. Thus in the usual DMRG language,
 the site $p$ canonical form corresponds to the DMRG wavefunction
 in the basis associated with  the  block configuration $\fbox{$\bullet_1 \ldots  \bullet_{p-1}$} \
\bullet_{p} \ \fbox{$\bullet_{p+1} \ldots  \bullet_{k}$}$.

A one-site DMRG wavefunction expressed in the canonical form
of a given site $p$ can always be expressed in the canonical form for any
other site (or using the traditional DMRG language, the DMRG
wavefunction for a given one-site block configuration can always be
expressed in the basis of any other one-site block configuration along a sweep). Since we are simply re-expressing 
the same wavefunction  in a different basis, the  coefficients $C$ and
site-functions $L,
R$ at different sites are related. 
To see the link explicitly, 
we compare the canonical forms at adjacent sites $p$, $p+1$ 
% For our  later algorithm to  evaluate of the two-particle density
% matrix, we will need to make use of canonical representations at
% \textit{different} sites, and certain computations which are difficult
% in one canonical representation become easy in another. 
% When changing
% between representations, we are simply re-expressing 
% the same wavefunction  in a different space, and this that the  wavefunction coefficients $C$ and
% transformation matrices $L,
% R$ at different sites are related. 
\begin{widetext}
\begin{align}
|\Psi \rangle& = \sum_{n_1 \ldots n_p \ldots n_k}
%%\bigl( 
{\boldL}^{n_1} \ldots \boldL^{n_{p-1}} 
%%\bigr)
%%\bigl( 
\boldC^{n_p} 
\boldR^{n_{p+1}}
%%\bigr)
%%\bigl( 
\boldR^{n_{p+2}}\ldots \boldR^{n_k}
%%\bigr) 
|n_1 \ldots n_p
 \ldots n_k\rangle\\
& = \sum_{n_1 \ldots n_p \ldots n_k}
%%\bigl(
\boldL^{n_1} \ldots \boldL^{n_{p-1}}
%%\bigr)
%%\bigl(
\boldL^{n_{p}}
 \boldC^{n_{p+1}}
%%\bigr)
%%\bigl(
\boldR^{n_{p+2}} \ldots \boldR^{n_k}
%%\bigr)
|n_1 \ldots n_p  \ldots n_k\rangle .
\end{align}
\end{widetext}
%% \begin{align}
%% |\Psi \rangle& = \mathop{\sum_{n_1 \ldots n_p \ldots n_k}}_{l_1
%%  \ldots l_{p-1}, r_p \ldots r_{k-1}} 
%% \Bigl( \boldL^{n_1}_{l_1} \ldots \boldL^{n_{p-1}}_{l_{p-2} l_{p-1}} \Bigr)
%% \Bigl( C^{n_p}_{l_{p-1} r_p} 
%% R^{n_{p+1}}_{r_p r_{p+1}}\Bigr)
%% \Bigl( R^{n_{p+2}}_{r_{p+1} r_{p+2}}\ldots R^{n_k}_{r_{k-1}}\Bigr) |n_1 \ldots n_p
%%  \ldots n_k\rangle\\
%% & = \mathop{\sum_{n_1 \ldots n_p \ldots n_k}}_{l_1
%%  \ldots l_{p-1}, r_p \ldots r_{k-1}} 
%% \Bigl(
%% \boldL^{n_1}_{l_1} \ldots \boldL^{n_{p-1}}_{l_{p-2} l_{p-1}} 
%% \Bigr)
%% \Bigl(
%% \boldL^{n_{p}}_{l_{p-1}
%%  l_{p}}
%%  C^{n_{p+1}}_{l_{p} r_{p+1}} 
%% \Bigr)
%% \Bigl(
%% R^{n_{p+2}}_{r_{p+1} r_{p+2}} \ldots R^{n_k}_{r_{k-1}} 
%% \Bigr)
%% |n_1 \ldots n_p  \ldots n_k\rangle
%% \end{align}
which yields the relation
\begin{align}
\boldL^{n_{p}} \boldC^{n_{p+1}}=\boldC^{n_p} \boldR^{n_{p+1}} \label{eq:wftrans}
\end{align}
% or, switching to the compound matrix view for
% $C^p, C^{p+1},L^p, R^{p+1}$
% \begin{align}
% \sum_{r} C^p_{ln,r} R^{p+1}_{r, r^\prime n} = \sum_{l^\prime} L^{p}_{ln, l^\prime}
% C^{p+1}_{l^\prime, n^\prime r^\prime} .
% %% \sum_{r_p} C^{n_p}_{l_{p-1}r_p} R^{n_p+1}_{r_pr_{p+1}}=\sum{l_p} L^{n_p}_{l_{p-1} l_p}
% %% C^{n_{p+1}}_{l_p r_{p+1}}
% \end{align}
Now say we are given $\boldC^{n_p} \boldR^{n_{p+1}}$ from the site
$p$ canonical form, and we wish to
determine $\boldL^{n_{p}} \boldC^{n_{p+1}}$ for the site $p+1$
canonical form, where $\boldL^{n_p}$  satisfies the orthogonality conditions
(\ref{eq:lorth}). We can obtain such a $\boldL^{n_p}$ solution  of (\ref{eq:wftrans}) together
with  $\boldC^{n_{p+1}}$ from the singular value decomposition (SVD) of
  $\boldC^{n_p}$,  viewed as the $4M \times M$ matrix with row
  indices $l_{p-1}n_p$, column indices $r_{p+1}$ and $M$ singular
  values $\sigma_{l_p}$,
\begin{align}
C^{n_p}_{l_{p-1},r_{p+1}} &= \sum_{l_p} L^{n_{p}}_{l_{p-1}, l_p} \sigma_{l_p}
V_{l_p r_{p+1}} \label{eq:density_basis},\\
C^{n_{p+1}}_{l_p,r_{p+2}} &= \sum_{ r_{p+1}} \sigma_{l_p} V_{l_p r_{p+1}}
R^{n_{p+1}}_{r_{p+1}, r_{p+2}} \label{eq:wfsvdtrans}
%% C^{n_p}_{l_{p-1}r_p}& = L^{n_p}_{l_{p-1} l_p} \sigma_{l_p} V_{l_p r_p}
%% R^{n_{p+1}}_{r_p r_{p+1}} \\
%% C^{n_p+1} &= \sum_{l_p r_p} \sigma_{l_p} V_{l_p r_p} R^{n_{p+1}}_{r_p r_{p+1}}
\end{align}
The above transformation between canonical forms at adjacent sites
corresponds directly to the transformation between block configurations
during  the sweep algorithm in the DMRG. In particular, Eq. (\ref{eq:density_basis}) corresponds to the
determination of the basis of the renormalised block
$\fbox{$\bullet_1 \ldots  \bullet_{p+1}$}$  from the density matrix eigenvectors of the superblock
 $\fbox{$\bullet_1 \ldots \bullet_p$} \ \bullet_{p+1}$, while Eq. (\ref{eq:wfsvdtrans}) corresponds to the
 wavefunction transformation  used to generate the guess at a given
 block configuration from that at the previous configuration. 
We note in passing that an \textit{exact} transformation between
canonical forms at different sites is only possible with the
\textit{one-site} DMRG ansatz. Most DMRG calculations use the two-site
DMRG ansatz with the block configuration $\fbox{$\bullet_1 \ldots  \bullet_{p-1}$}
\ \bullet_{p} \bullet_{p+1} \ \fbox{$\bullet_{p+2} \ldots  \bullet_{k}$}$
and a corresponding canonical  form at site $p$ 
\begin{widetext}
\begin{align}
|\Psi \rangle& = \sum_{n_1 \ldots n_p \ldots n_k} {\boldL}^{n_1} \ldots {\boldL}^{n_{p-1}}
 {\boldC}^{n_p n_{p+1}} {\boldR}^{n_{p+2}} \ldots {\boldR}^{n_k} |n_1
 \ldots n_p n_{p+1}
 \ldots n_k\rangle \\
& = \sum_{l_{p-1} n_p n_{p+1} r_{p+2}}
C^{n_p n_{p+1}}_{l_{p-1} r_{p+2}}
 |l_{p-1} n_p n_{p+1} r_{p+2}\rangle
\end{align}
\end{widetext}
% Here we see that two complete orbital Fock spaces $\{ n_p\}, \{n_{p+1}\}$
% appear in the wavefunction expansion, and consequently, 
Unlike in the one-site ansatz, the coefficient
matrix $\boldC^{n_p n_{p+1}}$ has a different shape from the $L$ and
$R$ site functions and has    $4M$ (as opposed to $M$ in the one-site case)
singular values. Thus it can only be approximately represented by the sum over
$M$ singular values in Eq. (\ref{eq:wfsvdtrans}), and the resulting truncation corresponds to ``discarding states'', in the
DMRG algorithm. The primary benefit of the two-site DMRG ansatz is greater robustness of convergence in the DMRG sweeps
 but  for the purposes of orbital optimisation,  the one-site DMRG ansatz  provides a single consistent DMRG wavefunction in
all canonical forms and block configurations and is to be preferred.
 
\subsubsection{Reduced density matrix evaluation}

Our task now is, given a DMRG wavefunction written explicitly as
(\ref{eq:dmrg_mps}) or equivalently in the renormalised expansion (\ref{eq:blockwf}), to
find an efficient algorithm to evaluate the one- and two-particle
density matrices. 
From the renormalised form we see that
we will need 
matrix representations of operators in each of the three spaces $\{l_{p-1}\}, \{ n_p \}, \{ r_{p+1} \}$,
i.e. matrix elements $\langle {l^{p-1}} | \hat{O} |
{l^{p-1}}^\prime \rangle$, $\langle n^{p} | \hat{O} |
{n^{p}}^\prime \rangle$, $\langle r^{p+1} | \hat{O} |
{r^{p+1}}^\prime \rangle$.  
Matrix representations  in the left and right spaces are in general of
dimension $M\times M$, since there are $M$ left and right
states. While the direct
evaluation of the one-particle density matrix would  require $k^2$
operator representations and thus
$O(M^2 k^2)$ storage (presenting no particular difficulties as the
memory requirement for the usual DMRG algorithm is also $O(M^2 k^2)$)
the two-particle density matrix would require $O(M^2 k^4)$ storage
which is prohibitively expensive. (It might appear that when solving
the Schr\"odinger equation,  the action  ${H}
|\Psi\rangle$ would also involve $O(k^4)$ operators and $O(M^2k^4)$
storage. However, there we do
not need the action of the operators $a^\dag_i a^\dag_j a_k a_l$ individually, but only the
total $\sum_{ijkl} v_{ijkl} a^\dag_i a^\dag_j a_k a_l$,  so we can
form intermediates where   operators are precontracted with two-electron
integrals to save memory, and the efficient arrangement of such
intermediates lies at the heart of the quantum chemical DMRG algorithm).

% To solve the Schr\"odinger equation, we need the action of the
% Hamiltonian,  $\hat{H} |\Psi\rangle$ which would  appear to require
% storage of ${O(k^4)}$ operators  $a^\dag_i a^\dag_j a_k
% a_l$ with a concomittant memory cost of $O(M^2 k^4)$.
% However, the memory cost of the standard quantum chemical DMRG
% algorithm is only $O(M^2 k^2)$. This is because 
% we do not actually need the action of the operators $a^\dag_i a^\dag_j a_k a_l$ individually, but only the
% sum $\sum_{ijkl} v_{ijkl} a^\dag_i a^\dag_j a_k a_l$, and thus we can
% manipulate groupings of operators that are precontracted with two-electron
% integrals to save memory. When this is done in an efficient manner,
% the total memory cost of the quantum chemical DMRG algorithm is only
% $O(M^2 k^2)$. 

% However, as discussed in the section \ref{sec:overview}, the
% evaluation of  the orbital gradient in the simplest orbital optimisation
% procedure explicitly requires the one- and two-particle density matrix
% elements. The evaluation of the one-particle density matrix presents
% no essential difficulties, but, 
% to obtain all the two-particle density matrix  elements $\langle a^\dag_i a^\dag_j a_k a_l
% \rangle$ at a \textit{given} block configuration would  indeed require  manipulating  $O(k^4)$ operators
% and their $M \times M$ matrix representations, with the associated
% $O(M^2 k^4)$ memory cost, which would be prohibitively expensive.
\begin{figure}[t]
\includegraphics[width=3.5in]{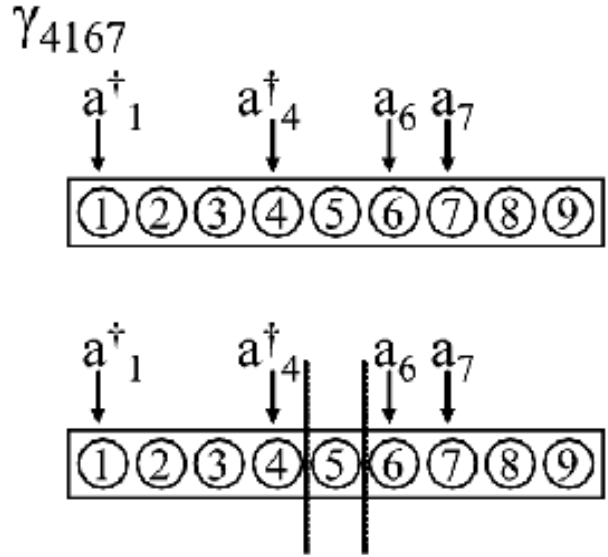}
\caption{\label{fig:dmop_split} Evaluation of a 2-rdm element
$\gamma_{4167}$. We can obtain this element e.g. at the block
configuration where indices $4,1$ are on the left block and indices
$6,7,$ are on the right block (corresponding to calling
$\textsc{Compute}(2,0,2)$ in Alg. \ref{algo:dm}).}
\end{figure}

The way forward is to observe that we are not tied to using a single
canonical form/block configuration for the DMRG wavefunction, but rather, can
evaluate a density matrix element  $\gamma_{ijkl}$ at any canonical form/block
configuration that is convenient. As we have described above, a given
DMRG wavefunction can be expressed in the canonical form/block-configuration associated with
any site. By  taking advantage of this flexibility, we can reduce the
memory requirements once again back to  $O(M^2 k^2)$, i.e. the same as
in the standard quantum chemical DMRG algorithm. 
Given  a two-particle density matrix element $\langle a^\dag_i a^\dag_j a_k
a_l\rangle$, where, say $i \leq j \leq k \leq l$, we choose a
block configuration such that $i, j$ lie in the left block  and
sites $k,l $  lie in the right block, i.e.  $\fbox{$ \ldots \bullet_{i} \ldots \bullet_{j}
  \ldots$} \ \bullet_p \ \fbox{$ \ldots \bullet_{k} \ldots
  \bullet_{l} \ldots$}$. The corresponding  matrix element may then
be evaluated  using  $a^\dag_i a^\dag_j$ on the left block, and $a_k
a_l$ on the right block, and thus no operator matrices with more than two
orbital indices appear on either block (see Figure \ref{fig:dmop_split}). 
By the appropriate choice of partitioning between the left and right
blocks,   we can arrange things such that we never manipulate  operators with more than two
orbital labels on either the left or right blocks for any
$ijkl$. During a DMRG sweep we iterate through all block
configurations where the dividing site $\bullet_p$ ranges from site 2
to site $k-1$. At each block configuration, we  then evaluate all
the two-particle density matrix elements which do not require more
than two-index operators on either the left or right blocks, 
 and assemble  the contributions of  all the block configurations  at
the end of  the DMRG sweep.

Along these lines, we can formulate an efficient
algorithm to evaluate the
two-particle density matrix  with a total per-sweep computational
cost of $O(M^3 k^4)$ and  a memory cost of $O(M^2 k^2)$. The
pseudocode is given in  Algs. (\ref{algo:dm}),
(\ref{algo:compute}). Alg. (\ref{algo:dm}) describes how to partition
the evaluation of different density matrix elements amongst the
block configurations as we traverse a DMRG sweep. The actual
calculation of the density matrix elements is carried out by the
function \textsc{Compute} in Alg. (\ref{algo:compute}), which computes
all density matrix elements that may be assembled from $nl$ index
operators on the left block, $np$ index operators on site $p$, and
$nr$ index operators on the right block.

\algsetup{indent=1em}
\newcommand{\Compute}{\ensuremath{\mbox{\sc Compute}}}
\newcommand{\Left}{\ensuremath{left}}
\newcommand{\SiteP}{\ensuremath{sitep}}
\newcommand{\Right}{\ensuremath{right}}
\begin{algorithm}
\caption{Two-particle density matrix evaluation showing how
  the two-particle density matrix is assembled across a DMRG sweep. \label{algo:dm}}
\begin{algorithmic}
\STATE{\textbf{special treatment for first configuration} $\fbox{$\bullet_1$} \ \bullet_2 \ \fbox{$\bullet_3 \ldots \bullet_k$}$}
\STATE{\Left $=$ site $1$, \SiteP $=$  site $2$, \Right $=$ sites $3
  \ldots k$}
\STATE{\Compute(4, 0, 0, \Left, \SiteP, \Right)}
\STATE{\Compute(3, 1, 0, \Left, \SiteP, \Right)}
\STATE{\Compute(3, 0, 1, \Left, \SiteP, \Right)}
\STATE{\Compute(2, 1, 1, \Left, \SiteP, \Right)}
\STATE{\textbf{sweep through block configurations} $\fbox{$\bullet_1 \ldots \bullet_{p-1}$} \ \bullet_p \ \fbox{$\bullet_{p+1} \ldots \bullet_k$}$}
\FOR{\SiteP  $=$ 2 to k-1}
  \STATE{\Left $=$ sites $1 \ldots p-1$, \Right $=$ sites $p+1 \ldots k$}
  \STATE{\Compute(1, 2, 1, \Left, \SiteP, \Right)}
  \STATE{\Compute(2, 1, 1, \Left, \SiteP, \Right)}
  \STATE{\Compute(2, 2, 0, \Left, \SiteP, \Right)}
  \STATE{\Compute(1, 3, 0, \Left, \SiteP, \Right)}
  \STATE{\Compute(0, 3, 1, \Left, \SiteP, \Right)}
  \STATE{\Compute(0, 4, 0, \Left, \SiteP, \Right)}
\ENDFOR
\STATE{\textbf{special treatment for final configuration} $\fbox{$\bullet_1 \ldots \bullet_{k-2}$} \ \bullet_{k-1} \ \fbox{$\bullet_k$}$}
\STATE{\Left $=$ sites $1\ldots k-2$, \SiteP $=$ site $k-1$, \Right $=$ site $k$}
\STATE{\Compute(0, 0, 4, \Left, \SiteP, \Right)}
\STATE{\Compute(0, 1, 3, \Left, \SiteP, \Right)}
\STATE{\Compute(1, 0, 3, \Left, \SiteP, \Right)}
\STATE{\Compute(0, 2, 2, \Left, \SiteP, \Right)}
\STATE{\Compute(2, 0, 2, \Left, \SiteP, \Right)}
\STATE{\Compute(1, 1, 2, \Left, \SiteP, \Right)}
\STATE{\Compute(1, 2, 1, \Left, \SiteP, \Right)}
\end{algorithmic}
\end{algorithm}

\newcommand{\opl}{\ensuremath{opl}}
\newcommand{\opp}{\ensuremath{opp}}
\newcommand{\opr}{\ensuremath{opr}}
\begin{algorithm}
\caption{ $\Compute(nl,np,nr,\Left,\SiteP,\Right)$. Note $nl, np, nr
  \leq 2$ and $nl+np+nr=4$, i.e. the number of indices in the two-particle
  density matrix $\gamma$. \label{algo:compute}}
\begin{algorithmic}
\FORALL{\opl $=$ operators with $nl$ indices on block $\Left$}
\STATE{\emph{(If parallel, loop only over $\opl$ stored on current proc)}}
\FORALL{\opp $=$ operators with $np$ indices on block \SiteP}
\FORALL{\opr $=$ operators with $nr$ indices on block \Right}
\STATE{$\gamma(np,nl,nr)$ $=$
  $\text{parity}(\opl,\opp,\opr) \times  \langle \Psi | \opl \otimes
  \opp \otimes \opr | \Psi \rangle$}
\ENDFOR
\ENDFOR
\ENDFOR
\STATE{\emph{(If parallel, accumulate contributions from all procs to root processor)}}
\end{algorithmic}
\end{algorithm}

An attractive  feature of the  quantum chemical DMRG algorithm is the
high level of parallelisability, which we have described in detail in
Ref. \cite{Chan2004}. In our implementation, the loops over 
operators in Alg. (\ref{algo:compute})  are trivially parallelised
because of how our  operators are divided across processors in our
original formulation \cite{Chan2004}. For example, the dominant computational cost of the two-particle
density matrix evaluation comes from  $\Compute(2, 1, 1, \Left,
\SiteP, \Right)$ in Alg. (\ref{algo:dm}), which costs $O(M^3 k^4)$ per DMRG sweep. However,
in our parallel DMRG implementation, the  two index operators $opl$ on the left block, namely $a^\dag_i a_j$ and
$a_i a_j$, are divided across the processors, while the
corresponding  one index
operators $opp, opr$ are replicated on all processors,   and thus we can easily
parallelise over the first $opl$ loop  in Alg. (\ref{algo:compute}). This
leads to a final  computational cost per sweep of $O(M^3 k^4/n_p)$
with a communication cost of $O(k^4 \ln n_p)$, where $n_p$ is the number of
processors.

\subsection{Orbital step and integral transformation}

\label{sec:orbital}
%\input{header.tex}
%\begin{document}

As described earlier, the DMRG wavefunction is  primarily efficient
 at capturing static correlation and consequently we   employ
 an active space DMRG description of the electronic structure, the
 purpose of the orbital optimisation then being to obtain the best
 form of the active space. 
Recall that the active space is defined by partitioning the orbitals into three
sets, closed-shell orbitals which remain doubly occupied in all DMRG
configurations, active orbitals which form the product active
space $\{ n_1 \} \otimes  \ldots \otimes \{ n_k\}$ in the DMRG
wavefunction expansion  (\ref{eq:dmrg_ansatz}), and external
orbitals, which remain unoccupied in all DMRG configurations. With
this partitioning, the active space DMRG wavefunction 
is determined with respect to  the active space Hamiltonian
\begin{equation}
H^\mathrm{act} = E^\mathrm{closed}
               + \sum_{ij} t^\mathrm{act}_{ij} a^\dag_i a_j
               + \sum_{ijkl} v_{ijkl} a^\dag_i a^\dag_j a_k a_l \label{eq:qc_Hact}
\end{equation}
where indices $i,j$ are limited to the active orbitals
and the modified one-particle integrals $t^\text{act}_{ij}$ and closed-shell
energy are given respectively by 
\begin{align}
E^\mathrm{closed} &= \sum_{c} t_{cc} + \sum_{cc'} (v_{cc'c'c} -
v_{cc'cc'}) \label{eq:closed_e}\\
t^\mathrm{act}_{ij} &= t_{ij} + 2 \sum_{c} (v_{iccj} - v_{icjc}) \label{eq:closed_t}
\end{align}
where $c, c^\prime$ denote the closed-shell indices.

% the method to describe correlation within an active space. First, we 
% partition the orbital space into three set of orbitals, 
% use the DMRG to describe correlation within an \textit{active} space
% of orbitals. 

Orbital optimisation chooses the best form of the active orbitals by 
  minimising the energy of the DMRG wavefunction with 
respect to the active and closed-shell orbitals. This is the basic idea behind the Complete-Active-Space Self-Consistent
Field (CASSCF) description of electronic structure. In CASSCF, the
active space wavefunction  is the exact eigenfunction
of the active space Hamiltonian (\ref{eq:qc_Hact}) and is thus invariant with respect to
active-active orbital rotations.   In the corresponding orbital
optimised DMRG-CASSCF, the accuracy of our active space DMRG wavefunction depends on the
 size of $M$, but  in this study we will use sufficiently large $M$ so that
 our wavefunction is nearly an exact eigenfunction of the active
 space Hamiltonian, and we will similarly omit active-active rotations.

The algorithm we use for  orbital optimisation is an Augmented Hessian
Newton Raphson scheme similar to that used in modern CASSCF
implementations \cite{knowles, yeager1982nra, lengsfieldiii1981som}. The orbital rotations are parameterised by the anti-hermitian
amplitudes $A$  in Eq. (\ref{eq:exp_rot}), and the derivative with respect
to these amplitudes is evaluated from the  one- and two-particle
density matrices from the   DMRG calculation. However, as the DMRG enables the use of 
larger active spaces than in traditional  CASSCF studies and
consequently we can expect to  have a larger number of
correlating external and closed-shell orbitals, we have focused on an
efficient parallel implementation of the
orbital optimisation. Here the primary task is to   parallelise the
four-index transformation  which is  performed after each orbital rotation to generate the
two-electron integrals in the basis of  the rotated orbitals. We now
describe how this is done.

Say we have a coefficient matrix $\boldsymbol{U}$ giving the expansion
coefficients for our rotated orbitals in terms of the starting atomic
orbitals. Then, the transformed integrals  $v_{pqrs}$ are obtained
from the atomic orbital integrals $v^\text{AO}_{\mu\nu\kappa\lambda}$
through (assuming real coefficients, for simplicity)
\begin{align}
v_{pqrs} = \sum_{\mu\nu\kappa\lambda} U_{p\mu} U_{q\nu} U_{r\kappa} U_{s\lambda} v^\mathrm{AO}_{\mu\nu\kappa\lambda}
\end{align}
As is well known, the four-index transformation should be carried out in four
quarter-transformation steps corresponding to the four contractions
with the coefficient matrices above. In our parallel transformation scheme, we
consider the four steps in two stages; in the first stage
we perform two quarter-transformations to construct 
half-transformed Coulomb and exchange intermediates $J, K$ 
% \begin{align} TY old
% [J_{ab}]_{pq} = \sum_{\mu\nu\kappa\lambda} C_{a\mu} C_{c\nu} C_{d\kappa} C_{b\lambda} v^\mathrm{AO}_{\mu\nu\kappa\lambda} \label{eq:tei_J}\\
% [K_{ab}]_{pq} = \sum_{\mu\nu\kappa\lambda} C_{a\mu} C_{c\nu} C_{b\kappa} C_{d\lambda} v^\mathrm{AO}_{\mu\nu\kappa\lambda} \label{eq:tei_K}
% \end{align}
% \begin{align} Garnet
% J_{ab}(\mu,\kappa) = \sum_{\nu\lambda} U_{a\nu}  U_{b\lambda} v^\mathrm{AO}_{\nu\mu\kappa\lambda} \label{eq:tei_J}\\
% K_{ab}(\mu,\lambda) = \sum_{\nu\kappa} U_{a\nu}  U_{b\kappa} v^\mathrm{AO}_{\nu\mu\kappa\lambda} \label{eq:tei_K}
% \end{align}
\begin{align}
J_{ab}(\nu,\kappa) = \sum_{\mu\lambda} U_{a\mu}  U_{b\lambda} v^\mathrm{AO}_{\mu\nu\kappa\lambda} \label{eq:tei_J_1}\\
K_{ab}(\nu,\kappa) = \sum_{\mu\lambda} U_{a\mu}  U_{b\lambda} v^\mathrm{AO}_{\mu\nu\lambda\kappa} \label{eq:tei_K_1}
\end{align}
while in the second stage, we perform the remaining quarter
transformations on the $J$, $K$ intermediates to obtain the final integrals
%\begin{align}
%v_{acdb}& = \sum_{\mu\kappa} J_{ab}(\mu,\kappa )  U_{c\mu} U_{d\kappa}\\
%v_{acbd}& = \sum_{\mu\kappa} K_{ab}(\mu,\kappa )  U_{c\mu} U_{d\kappa}
%\end{align}
\begin{align}
[J_{ab}]_{pq} &= v_{apqb} = \sum_{\nu\kappa} J_{ab}(\nu,\kappa ) U_{p\nu} U_{q\kappa} \label{eq:tei_J_2} \\
[K_{ab}]_{pq} &= v_{apbq} = \sum_{\nu\kappa} K_{ab}(\nu,\kappa ) U_{p\nu} U_{q\kappa} \label{eq:tei_K_2}
\end{align}
Note that for the purposes of optimising the active orbitals, we only
need the integrals that appear in the augmented Hessian. Thus, the $ab$ indices in
(\ref{eq:tei_J_1}), (\ref{eq:tei_K_1}) only need to run over the active orbitals  while the $pq$
indices  need to run over all the closed-shell, active, and external orbitals.

In the first stage, we parallelise the construction of the  $J, K$ intermediates by 
dividing up the intermediates according to their untransformed AO
indices. For example, the construction of $J_{ab}(\nu,\kappa)$ is divided amongst
the processors according to the pair of indices $(\nu,\kappa)$;  each processor is
then responsible for constructing the $J$ intermediates for all $(\bar{\nu},\bar{\kappa}) \in
\text{proc}$. This allows us to also partition the AO integrals
amongst the processors according to the same divided pair of indices ($\bar{\nu},\bar{\kappa}$); e.g. to
construct $J_{ab}(\bar{\nu},\bar{\kappa})$ for $(\bar{\nu},\bar{\kappa}) \in \text{proc}$ we only
need AO integrals such as $v^\text{AO}_{\mu\bar{\nu}\bar{\kappa}\lambda}$ for $(\bar{\nu},\bar{\kappa}) \in
\text{proc}$ to be stored on that  processor.

Once all $J$ and $K$ intermediates are constructed, we parallelise the
second stage  with respect to the transformed $ab$ indices of the $J$,
$K$ intermediates. Thus $ab$ is divided amongst the processors, and
each processor constructs the final integrals $v_{\bar{a}pq\bar{b}}, v_{\bar{a}p\bar{b}q}$ for all
$\{\bar{a}\bar{b}\} \in \text{proc}$. Since  the first stage is parallelised over
a pair of AO indices ($\nu,\kappa$) (and the $J$ and $K$ intermediates are
divided across the processors accordingly) while the second stage is parallelised  over the
two transformed indices ($ab$), we need to redistribute  the intermediates $J$
and $K$ amongst the processors  between the first and
second stages. This is the main communication step.

In addition to  above parallelisation, further
efficiencies can be gained by using the permutational and spatial
symmetries of the integrals. Our complete parallelised algorithm, which uses these symmetries, is
presented in pseudocode in Alg. (\ref{alg:para_itrf}).
The cost of the four-index integral transformation as implemented is
$O((K^4k+K^3k^2)/n_{p})$  for CPU, $O((K^4 +
K^2k^2)/n_{p})$  for disk space, $O(K^2k^2/n_{p})$  for memory,
and $O(K^2k^2)$ for overall communication, where $K$ is the total
number of orbitals, $k$ is the number of active orbitals, and $n_{p}$
is the number of processors.

To complete our efficient implementation of orbital optimisation, we
have also parallelised the remaining steps in the Augmented Hessian Newton-Raphson
solver. These additional steps take up only a small part of the
computational time and have an overall cost $O(K^2k^3/n_{p})$  for CPU time,
$O(K^2k^2/n_{p})$  for memory,
$O(Kk)$ for  communication.

%  construct the
% $J$ and $K$ integrals in such a way that the pairs of the
% indices $\{ab\}$ that are divided up by $n_{nproc}$ are
% assigned to the different processors, which eventually own $J_{ab}$
% and $K_{ab}$.  For the construction, the AO integrals
% $v^\mathrm{AO}_{\mu\nu\kappa\lambda}$ are distributed across the
% processors.

% which constructs the  new two-electron integrals (TEIs) after each
% orbital rotation.
% into the CAS orbital
% basis, which may be of the localised orbital form or natural orbital
% form.  The TEIs for the input of DMRG calculations are obtained by
% transoforming the TEIs from the AO basis to
% the active orbital basis of the localised orbital form,
% \begin{equation}
% v_{ijkl} = \sum_{\mu\nu\kappa\lambda} C_{i\mu} C_{j\nu} C_{k\kappa} C_{l\lambda} v^\mathrm{AO}_{\mu\nu\kappa\lambda}
% \end{equation}

% Our primal effort to parallelize the orbital optimisation step is
% placed on an efficient parallelization of 

{
\renewcommand{\baselinestretch}{1.2}
\begin{algorithm}[H]
\caption{Parallel four-index integral transformation algorithm.}
\label{alg:para_itrf}
\begin{algorithmic}
\STATE{\textbf{Stage 1: Assemble $J$ and $K$ intermediates}}
\STATE{Divide AO integrals $v^\text{AO}_{\mu\nu\kappa\lambda}$ by a factor $(2-\delta_{\mu\lambda})(2-\delta_{\nu\kappa})(2-\delta_{\mu\lambda,\nu\kappa})$}
\FOR{$\bar{\nu}, \bar{\kappa} \, (\bar{\nu} \ge \bar{\kappa}) \in \text{proc}$}
  \FOR{$a, \mu, \lambda$ $\,$ s.t. $\mu \ge \lambda$, $\mu\lambda \ge \bar{\nu}\bar{\kappa}$ }
  \STATE{$M^a_\mu    (\bar{\nu}, \bar{\kappa})$ {\tt +=} $v^\mathrm{AO}_{\mu\bar{\nu}\bar{\kappa}\lambda} \, U_{a\lambda}$; \,\,\,
         $N^a_\lambda(\bar{\nu}, \bar{\kappa})$ {\tt +=} $v^\mathrm{AO}_{\mu\bar{\nu}\bar{\kappa}\lambda} \, U_{a\mu    }$}
  \STATE{$N^a_\mu    (\bar{\nu}, \bar{\kappa})$ {\tt +=} $v^\mathrm{AO}_{\mu\bar{\kappa}\bar{\nu}\lambda} \, U_{a\lambda}$; \,\,\,
         $N^a_\lambda(\bar{\nu}, \bar{\kappa})$ {\tt +=} $v^\mathrm{AO}_{\mu\bar{\kappa}\bar{\nu}\lambda} \, U_{a\mu    }$}
  \ENDFOR
  \FOR{$a, \lambda$}
  \STATE{$N^a_\lambda(\bar{\nu}, \bar{\kappa})$ {\tt +=} $M^a_\lambda(\bar{\nu}, \bar{\kappa})$}
  \ENDFOR
  \FOR{$a, \mu, \lambda$ $\,$ s.t. $\mu \ge \lambda$, $\bar{\nu}\bar{\kappa} \ge \mu\lambda$ }
  \STATE{$L^a_\mu    (\bar{\nu}, \bar{\kappa})$ {\tt +=} $v^\mathrm{AO}_{\bar{\nu}\mu\lambda\bar{\kappa}} \, U_{a\lambda}$}
  \ENDFOR
  \FOR{$a,b,\lambda$ $\,$ s.t. $a \ge b$}
  \STATE{$J_{ab}(\bar{\nu}, \bar{\kappa})$ {\tt +=} $M^a_\lambda(\bar{\nu}, \bar{\kappa}) \, U_{b\lambda} 
                                                   + M^b_\lambda(\bar{\nu}, \bar{\kappa}) \, U_{a\lambda} 
                                                   + L^a_\lambda(\bar{\nu}, \bar{\kappa}) \, U_{b\lambda} 
                                                   + L^b_\lambda(\bar{\nu}, \bar{\kappa}) \, U_{a\lambda}$}
  \ENDFOR
  \FOR{$a,b,\lambda$}
  \STATE{$K_{ab}(\bar{\nu}, \lambda)$ {\tt +=} $N^a_\lambda(\bar{\nu}, \bar{\kappa}) \, U_{b\bar{\kappa}}$}
  \ENDFOR
\ENDFOR
\FOR{$a,b$ $\,$ s.t. $a \ge b$}
\STATE{write $J_{ab}$, $K_{ab}$, and $K_{ba}$ on disk}
\ENDFOR
\STATE{\textbf{Stage 2: Redistribute $J$ and $K$, transform to final integrals}}
\FOR{$a, b \, (a \ge b)$}
\STATE{read $J_{ab}$, $K_{ab}$, $K_{ba}$ from disk and send to proc$(a,b)$}
\ENDFOR
\FOR{$\bar{a}, \bar{b} \, (\bar{a} \ge \bar{b}) \in $ proc, $\nu, \kappa \, (\nu \ge \kappa)$}
\STATE{$J_{\bar{a}\bar{b}}(\kappa,\nu) += J_{\bar{a}\bar{b}}(\nu,\kappa)$}
\ENDFOR
\FOR{$\bar{a}, \bar{b} \, (\bar{a} \ge \bar{b}) \in $ proc, $\nu, \kappa $}
\STATE{$K_{\bar{a}\bar{b}}(\kappa,\nu) += K_{\bar{b}\bar{a}}(\nu,\kappa)$}
\ENDFOR
\FOR{$\bar{a}, \bar{b} \, (\bar{a} \ge \bar{b}) \in $ proc, $p,q,\nu,\kappa$}
\STATE{$v_{\bar{a}pq\bar{b}}$ {\tt +=} $J_{\bar{a}\bar{b}}(\nu,\kappa) \, U_{p\nu} U_{q\kappa}$ (eqn.\ (\ref{eq:tei_J_2}))} 
\STATE{$v_{\bar{a}p\bar{b}q}$ {\tt +=} $K_{\bar{a}\bar{b}}(\nu,\kappa) \, U_{p\nu} U_{q\kappa}$ (eqn.\ (\ref{eq:tei_K_2}))} 
\ENDFOR
%\ENDFOR
\end{algorithmic}
\end{algorithm}
}

\subsection{Complete Orbital Optimised DMRG-CASSCF Algorithm}

\label{sec:complete}
With the description of the density matrix evaluation in
Sec. \ref{sec:structure} and the orbital optimisation and integral transformation in
Sec. \ref{sec:orbital}, we now have the basic ingredients to
perform the  DMRG-CASSCF algorithm,
according to the general outline in Sec. \ref{sec:overview}. 

There is one final ingredient however, the secret ingredient. 
As the DMRG works best in a
localised basis (particularly in larger systems) it is beneficial to
localise the active space after each orbital optimisation. 
We have done this using the Pipek-Mezey procedure \cite{pipekmezey}; the active-space integrals
are first transformed into this local basis before being input into
the DMRG calculation. In total therefore, 
the complete DMRG-CASSCF algorithm is as follows:
\begin{enumerate}
\item Localise the active space orbitals.
\item Transform the AO integrals to the active space basis and build the
  active space Hamiltonian.
\item Perform the DMRG calculation using the active space Hamiltonian.
\item From the converged DMRG wavefunctions at each block
  configuration, assemble the one- and two-particle density matrices.
\item Using the density matrices, obtain the orbital gradient and orbital step from the Augmented
  Hessian Newton-Raphson solver.
\item From the orbital step, determine  the new active space orbitals.
\item Goto 1. until convergence in the energy.
\end{enumerate}
Steps 1.-6. constitute a single DMRG-CASSCF \textit{macro-iteration}.

 \section{\label{sec:level4}Applications}
\label{sec:apps}

\subsection{Long Polyenes}
\label{sec:polyenes}

\subsubsection{Background}

Polyenes are the simplest conjugated systems, consisting of
alternating singly and doubly bonded carbons arranged in  a
chain. They are valuable models not only to understand
conjugated polymers of materials interest (e.g. poly-acetylene is
simply an infinite polyene)  but also biological molecules such as the
carotenoid and retinal families of pigments involved in photosynthesis and vision.
In these systems, the functionality of the molecules relies on the low-lying $\pi$-$\pi^*$ excited states of the
conjugated backbone, which serve as the conduits for energy  transfer.
 The excited states are labelled by their symmetry
under the $C_{2h}$ point group, giving rise to  $A_g, B_g, A_u, B_u$
symmetry labels. Furthermore, they are usually given an additional
$+/-$ label to indicate their approximate particle-hole symmetry. In
Hamiltonians (such as the H\"uckel Hamiltonian) which support symmetric
sets of energy states around the Fermi level, there is an additional
symmetry associated with rotating the molecular orbital diagram so
that the bonding and anti-bonding levels swap places \cite{pariser}.  Although
particle-hole symmetry is not a true symmetry of the \textit{ab-initio}
electronic Hamiltonian, it is still customary to use such labels for
the polyenes, in particular, because the $+/-$ states have very
different qualitative electronic structure;  valence bond
studies of the Hubbard model \cite{hubbardmodel} show that the $+$ states consist mainly of ionic
valence bond structures, while the $-$ states consist mainly
of covalent valence bond structures \cite{kurashige,Ramasesha1996,tavan}.

In this study we have looked only at singlet states and henceforth we
shall be considering singlet states only. 
The  ground state of the polyenes is known to always be of $A_g^-$
symmetry. The lowest dipole-allowed singlet transition, which has a predominantly
HOMO$\to$LUMO excitation character, has  $B_u^+$ symmetry. However,
contrary to what one might expect, this $1A_g^- 
\rightarrow 1B_u^+$ transition is not  the lowest singlet  transition \cite{kohler,kohler1}. 
Rather, as shown by Kohler \etal in octa-tetraene \cite{kohler}, there is a \textit{lower} dipole
forbidden excitation,  later identified as the $2A_g^-$ state, which can
be rationalised in valence bond language as arising from a pair of
singlet-triplet excitations in the two separate double
bonds that  recouple to form a singlet state \cite{andres,dunningshavitt,cave,cavedavidson,brooks,petrongolo,lappe,lasaga,bachler}. 
Following the
observation of the $2A_g^-$ state in octa-tetraene, there has been much debate over the correct
ordering of the $2A_g^-$ and $1B_u^+$ excited states in the shorter
polyenes, compounded both by experimental difficulties in observing
the dipole-forbidden $2A_g^-$ state as well as theoretical challenges
in achieving a balanced description of the two states, which are dominated
by very different kinds of correlation, namely static correlation in
the $2A_g^-$ state and dynamic correlation in the $1B_u^+$ state.
In longer polyenes and the biologically active carotenoid and retinal pigments, questions about the low-lying spectrum are not
restricted simply to the $2A_g^-$ and $1B_u^+$ state ordering. Recent
 studies using Resonance Raman excitation profiles (RREP) and
 electronic absorption spectroscopy on substituted polyenes in the
 carotenoid family, have indicated the presence of additional dark
states below the $1B_u^+$ state \cite{sashima,sashima2,fujii,onaka,furuichi}. In particular, for
the all-\textit{trans}-carotenoids with  (the number of double
bonds) $n= 9-11$, Sashima \etal observed a $1B_u^-$ state
between the $2A_g$ and $1B_u^+$ \cite{sashima, cogdellscience}. 
More recently, Furuichi \etal observed
a $3A_g^-$ level between the $1B_u^-$ and $1B_u^+$ states in
carotenoids with $n=11-13$, and assigned the tentative state ordering of
$1A_g^- < 2A_g^- < 1B_u^- < 3A_g^- < 1B_u^+$ \cite{furuichi}. The  assignment was made by
extrapolating from the earlier PPP-MRDCI calculations by Tavan and
Schulten on short polyenes ($n=2-8$), which had predicted the
existence of these additional states \cite{tavan}.

To better understand the electronic structure of these low-lying states,
we would ideally like to be able to carry out an \textit{ab-initio}
multireference calculation, using the complete $\pi$-valence space. 
However, the large number of active $\pi$ orbitals in the longer
polyenes means that it is not possible to
perform such calculations with traditional CAS algorithms 
for these systems.
Hirao and coworkers \cite{hirao, kurashige} carried out \textit{incomplete valence}
  CASSCF and CASCI-MRMP using a (10,10) active space on the polyene
series up to $\text{C}_{28}\text{H}_{30}$ and observed reasonable
agreement with experiment. However, with  our new orbital optimised DMRG-CASSCF procedure, we can now re-examine
 the low-lying excitations in these systems correlating the \textit{complete} $\pi$-
valence space even for the longer polyenes and carotenoids.

\subsubsection{Computational details}
\label{sec:polyene_compute}
The polyene molecular geometries for $\text{C}_8\text{H}_{10}, \text{C}_{12}\text{H}_{14},
\text{C}_{16}\text{H}_{18},  \text{C}_{20}\text{H}_{22}, \text{C}_{24}\text{H}_{26}$
were optimised at the density functional level  using the B3LYP functional \cite{becke1993dft,lee1988dcs} as implemented in
 \textsc{Gaussian03} \cite{gaussian}. The polyene molecules were constrained to have $C_{2h}$ symmetry, with
the $C_2$ axis as the $z$-axis. The cc-pVDZ basis \cite{cc-pvdz}  was used for all  calculations.

In our DMRG-CASSCF calculations we used a complete $\pi$-valence
space i.e. in $\text{C}_{24}\text{H}_{26}$, this was a
(24, 24) active space. 
To generate this active  space, we first performed a restricted Hartree-Fock
calculation in PSI3 \cite{PSI3,crawford2007sp} to obtain canonical Hartree-Fock molecular orbitals. From
these molecular orbitals, we could not trivially identify appropriate
$\pi$  anti-bonding active orbitals because of significant $2p$-$3p$
mixing. We constructed the anti-bonding component of the active space as a set of projected atomic
orbitals, by first projecting out the $\pi$
bonding space from a set of $2p_z$ atomic orbitals. These projected
atomic orbitals were then symmetrically orthogonalised, then
relocalised together with the bonding
molecular orbitals (using the Pipek-Mezey procedure \cite{pipekmezey})  to yield the
complete active space in our calculations. The final set of active
orbitals generated in this way resemble an orthogonal set of $2p_z$
orbitals. 

Note that our initial active space does not correspond
precisely to an active space obtained by selecting Hartree-Fock
canonical orbitals. Thus DMRG energies obtained \textit{before} orbital
optimisation do not correspond to typical CASCI energies,
but instead to CASCI energies obtained in our projected-atomic orbital
(PAO) virtual space. This distinction is noted in our tables with the 
abbreviation DMRG-PAO-CASCI. After orbital optimisation, however, our
DMRG-CASSCF energies do correspond to true CASSCF energies, up to the
accuracy of the DMRG calculation.

%  active
% space in the following fashion. The occupied part of the active space is chosen by simply looking at the pz character and
% thereby taking all the n/2 occupied active space which has appreciable 2pz character.
%  The virtual active space is chosen
% by projecting the already chosen occupied active space from 2pz atomic orbitals on the conjugated planar C atoms.
% This gives an active space that is predominantly 2pz in character. 
% The virtual space is then found by projecting out the virtual active space
% from all the virtual molecular orbitals (including the active space). After projection the
% redundancies are got rid of by choosing the first n orbitals. The molecular orbitals thus created are
% then orthogonalized by Gram-Schmidt and then the active space is localised using Pipek-Mezey scheme
% of localisation.

\begin{figure}
\centering
    \includegraphics[width=3.5in]{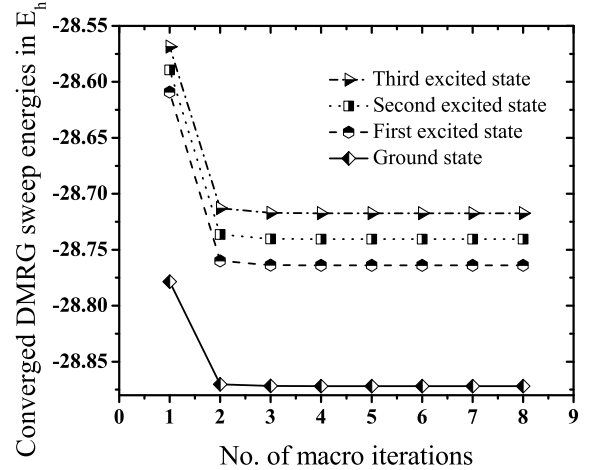}
\caption{\label{fig:dmrg} Converged DMRG sweep energies in Hartrees 
vs number of orbital optimisation macro iterations in $\text{C}_{20}\text{H}_{22}$.}
\end{figure}

We carried out state-averaged DMRG-CASSCF calculations  in the above active
space with the one-site DMRG algorithm with $M=250$  and
averaging over the $4$ lowest eigenstates.
The DMRG sweeps were  converged to $10^{-10}E_h$ in the DMRG
energy, which took  roughly 30 DMRG sweeps. The number of renormalised
states  was increased smoothly from a starting value of $M=50$ to the final value
of $M=250$. To aid the convergence of the
DMRG sweeps in the one-site algorithm, we applied a system-environment perturbation as described
in Ref. \cite{white-onedot}, with a starting magnitude of $10^{-3}$ that smoothly decreased
to 0 after 20 sweeps. We estimate the remaining error in the DMRG energies at
the $M=250$ level  from the exact Full-Configuration Interaction
energies  in the same active space to be less than  $0.1$m$E_h$.  
Our DMRG calculations were combined with orbital rotation in a  macro-iteration consisting of 
a converged DMRG calculation, an Augmented-Hessian step
based orbital rotation, integral transformation,  and orbital
localisation, as described in Sec. \ref{sec:complete}.  Typically 10-15 macro-iterations
of the complete DMRG/orbital optimisation cycle were necessary
to converge the energies to a tolerance of better than $10^{-6}
E_h$. The convergence of the state energies with the number of macro-iterations
is shown in Fig. \ref{fig:dmrg}.

% Up to 30 DMRG sweeps were performed and the 
% [Details on the DMRG procedure]

% The CASSCF calculation is done with this active space which is the localised molecular
% orbitals that are predominantly 2pz in character. For a typical calculation the DMRG takes about 30
% sweeps to converge and there are about 10-15 orbital rotations before we can self-consistently solve 
% for the active space.

The spatial and spin symmetries of excited states were assigned as
follows. Firstly, all  excited states were restricted to be of singlet
spin symmetry through the application of a shift $\lambda(\hat{S}^2-
\langle S\rangle(\langle S\rangle+1)$ with $\lambda=0.5$ \cite{Moritz2005rel}.
To obtain the spatial symmetry, the ground state was assumed to be
$1A_g^-$ as established by prior experimental and theoretical work. To
determine whether  the excited states were of $A_g$ or $B_u$ symmetry the transition dipole  matrices
were calculated between the  states.  Additionally, to determine the
approximate particle-hole $+$ or $-$  symmetry we examined the magnitude of the
transition dipoles;  large transition dipoles for an allowed
transition indicated that the transition involved a change of
particle-hole symmetry between the states.

% $A_g^-$ and excited $B_u$ state indicates $B_u^+$ symmetry, while a small
% dipole indicated $B_u^-$; a large transition dipole between the
% $B_u^+$ and excited $A_g$ state indicates $A_g^-$, while a small
% dipole indicates $A_g^{+}$. 

% or the particle hole symmetry (cite pariser), we look at the magnitude of the 
% dipole or the dipole element mentioned above.
% For the symmetry to be + the dipole element must be appreciable, i.e., there should be a
% considerable dipole moment and so an ionic state. On the other hand if it is small, we assign -
% symmetry or covalent state.

% We additionally calculated  the
% transition dipole matrix between the ground and excited
% states. 

% Assuming a $A_g^{-}$ symmetry for the ground state, we could
% assign the excited states by the size of the dipole matrix
% elements. Vanishing dipole matrix elements indicate $A_g$ 

% [properties, assigning states]
% The  is calculated to get the spatial symmetry of the excited states.
% The spin symmetry is already calculated from the expectation value of $S^2$. And by application
% of a projector only the singlet energies have been calculated. 
% $ \langle\Psi_0|n_j^{\alpha}+n_j^{\beta}-n_{C(j)}^{\alpha}-n_{C(j)}^{\beta}|\Psi_i\rangle$ is considered.
% C(j) is the index that is related to j by a $C_2$ rotation.
% The two values possible are 0 or non-zero. If it zero the transition is dipole forbidden and
% since the ground state is $A_g^-$, that excited state should have $A_g^-$ symmetry. If it is non-zero
% it is $B_u$ in symmetry. Now, 

\subsubsection{Discussion}

\begin{figure}
\centering
    \includegraphics[width=3.5in]{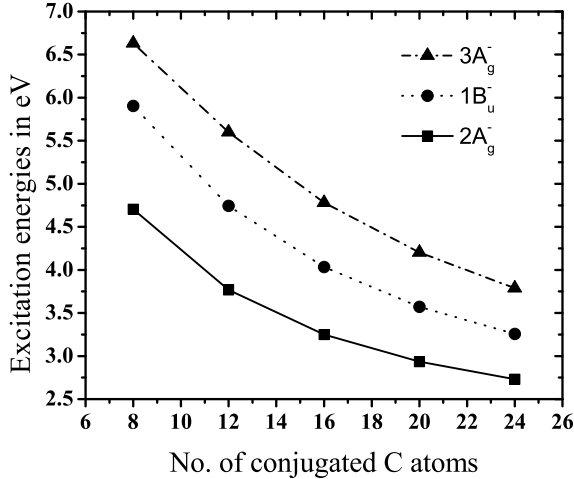}
\caption{\label{fig:casscf} DMRG-CASSCF excitation energies in $eV$ for the 
$2A_g^-$, $1B_u^-$ and $3A_g^-$ states in the conjugated polyenes 
$\text{C}_8\text{H}_{10}$ to $\text{C}_{24}\text{H}_{26}$ .}
\end{figure}

%% \begin{figure}[t1]
%% \centering
%%     \includegraphics[width=10cm]{casci.eps}
%% \caption{\label{fig:casci} CASCI}
%% \end{figure}

\begin{figure}
\centering
    \includegraphics[width=3.5in]{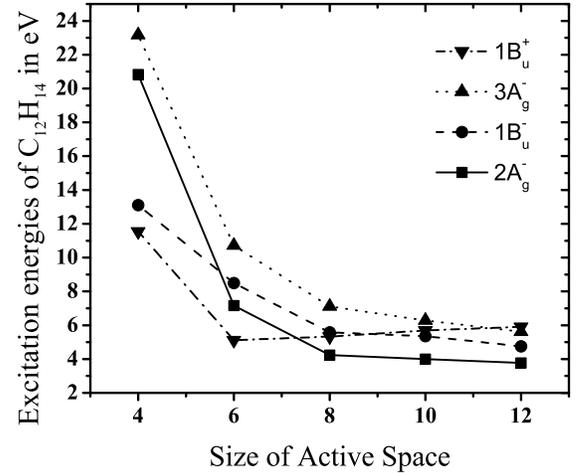}
\caption{\label{fig:activespace} Change in CASSCF energies of the low-lying states of
$\text{C}_{12}\text{H}_{14}$ as a function of increasing the active space from (4,4) to (12,12) (i.e. complete valence active space).}
\end{figure}

%% \begin{figure}[t1]
%% \centering
%%     \includegraphics[width=10cm]{pic/effect_of_or.eps}
%% \caption{\label{fig:or} Effect of Orbital Rotation}
%% \end{figure}

% \begin{figure}[htbp]
%   \begin{center}
%     \mbox{
%       \subfigure[1]
%                 {
%                   \includegraphics[width=5cm]{pic/no1.eps}
%                 } \quad
%       \subfigure[2]
%                 {
%                   \includegraphics[width=5cm]{pic/no2.eps}
%                 }
%       }
%     \mbox{
%       \subfigure[3]
%                 {
%                   \includegraphics[width=5cm]{pic/no3.eps}
%                 } \quad
%       \subfigure[4]
%                 {
%                   \includegraphics[width=5cm]{pic/no4.eps}
%                 }
%       }
%     \mbox{
%       \subfigure[5]
%                 {
%                   \includegraphics[width=5cm]{pic/no5.eps}
%                 } \quad
%       \subfigure[6]
%                 {
%                   \includegraphics[width=5cm]{pic/no6.eps}
%                 }
%       }
%     \mbox{
%       \subfigure[7]
%                 {
%                   \includegraphics[width=5cm]{pic/no7.eps}
%                 } \quad
%       \subfigure[8]
%                 {
%                   \includegraphics[width=5cm]{pic/no8.eps}
%                 }
%       }
%     \caption{\label{fig:no7} Natural Orbitals }
%   \end{center}
% \end{figure}

% \begin{figure}[t1]
% \centering
%     \includegraphics[width=7cm]{pic/LMO1.eps}
% \caption{\label{fig:lmo} Localised Orbitals}
% \end{figure}

\begin{table*}
  \centering
  \caption{\label{tab:polyene_energy}Energies, symmetries, and oscillator strengths for the lowest lying singlet excited states in conjugated polyenes.
The DMRG-PAO-CASCI and DMRG-CASSCF entries for the $1A_g^-$ ground-states  give the total energy in $E_h$; the other entries give the excitation
energies from the ground state in $eV$. The estimated error of the
  DMRG-CASSCF energies from the exact CASSCF energies in the same
  active space is less than 0.1m$E_h$. The notation $(n,m)$ denotes the active space used in the DMRG-PAO-CASCI and DMRG-CASSCF calculations. Oscillator strengths are in $a.u.$ for the ground-state, excited state transition. The CASCI-MRMP excitation energies are from 
Kurashige \etal \cite{kurashige}; note that these used at most a (10,10) active space. The experimental numbers in brackets are from  measurements on the substituted polyene, spheroidene \cite{furuichi}.}
  \begin{tabular}{|c|c|c|c|c|c|c|}
  \hline\hline  Polyenes        & Symmetry &  DMRG          & DMRG            &  Oscillator & CASCI-MRMP & Expt \\
                                &          &  PAO-CASCI     & CASSCF          &  Strength   &            &      \\
  \hline \hline                            
  $\text{C}_{8}\text{H}_{10}  $ & $1A_g^-$ &  $-308.823021$ & $-308.825879$   &             &        & \\
  $ (8,8)       $               & $2A_g^-$ &  $6.33       $ & $4.69       $   &  Forbidden  & $4.26$ & $3.54$ \footnotemark[1]\\
  $             $               & $1B_u^-$ &  $7.49       $ & $5.88       $   &  $0.0565$   & $5.30$ & \\
  $             $               & $3A_g^-$ &  $7.95       $ & $6.60       $   &  Forbidden  & $7.20$ & \\ \hline
  $\text{C}_{12}\text{H}_{14} $ & $1A_g^-$ &  $-462.661260$ & $-462.670591$   &             &        & \\
  $ (12,12)     $               & $2A_g^-$ &  $5.40       $ & $3.76       $   &  Forbidden  & $3.19$ & \\
  $             $               & $1B_u^-$ &  $6.30       $ & $4.74       $   &  $0.0620$   & $3.98$ & \\
  $             $               & $3A_g^-$ &  $7.01       $ & $5.59       $   &  Forbidden  & $5.12$ & \\ \hline
  $\text{C}_{16}\text{H}_{18} $ & $1A_g^-$ &  $-616.499262$ & $-616.514639$   &             &        & \\
  $ (16,16)     $               & $2A_g^-$ &  $4.90       $ & $3.25       $   &  Forbidden  & $2.50$ & $2.21$ \footnotemark[2]\\
  $             $               & $1B_u^-$ &  $5.60       $ & $4.03       $   &  $0.0502$   & $3.10$ & \\
  $             $               & $3A_g^-$ &  $6.28       $ & $4.78       $   &  Forbidden  & $3.99$ & \\ \hline
  $\text{C}_{20}\text{H}_{22} $ & $1A_g^-$ &  $-770.337112$ & $-770.358327$   &             &        & \\
  $ (20,20)     $               & $2A_g^-$ &  $4.60       $ & $2.93       $   &  Forbidden  & $2.04$ & $(1.76)$ \footnotemark[3]\\
  $             $               & $1B_u^-$ &  $5.15       $ & $3.57       $   &  $0.0427$   & $2.51$ & $(2.18)$ \footnotemark[3]\\
  $             $               & $3A_g^-$ &  $5.71       $ & $4.20       $   &  Forbidden  & $3.11$ & $(2.47)$\footnotemark[3]\\ \hline
  $\text{C}_{24}\text{H}_{26} $ & $1A_g^-$ &  $-924.174795$ & $-924.201821$   &             &        & \\
  $ (24,24)     $               & $2A_g^-$ &  $4.42       $ & $2.73       $   &  Forbidden  & $1.70$ & $(1.53)$\footnotemark[3]\\
  $             $               & $1B_u^-$ &  $4.85       $ & $3.25       $   &  $0.0384$   & $2.05$ & $(1.80)$\footnotemark[3]\\
  $             $               & $3A_g^-$ &  $5.31       $ & $3.78       $   &  Forbidden  & $2.45$ & $(2.02)$\footnotemark[3]\\ \hline

\hline\hline
\end{tabular}
\footnotetext[1]{\cite{granville}.}
\footnotetext[2]{\cite{kohlerc16}.}
\footnotetext[3]{\cite{furuichi}.}
\end{table*}

In Table \ref{tab:polyene_energy}  we present the energies, symmetries, and oscillator
strengths for the ground state
and first 3 excitations in the polyenes from
$\text{C}_8\text{H}_{10}$ to $\text{C}_{24}\text{H}_{26}$. For
comparison, we also give the excitation energies obtained from the CASCI-MRMP
calculations of Kurashige \etal \cite{kurashige}, as well as the experimental
energies where available. (Note that in $\text{C}_{20}\text{H}_{22}$,
the experimental excitation energies were obtained from the carotenoid spheroidene,
which has a $\text{C}_{20}$ conjugated backbone). 

We see that while our complete $\pi$-valence active space DMRG-CASSCF calculations generally
overestimate the excitation energies, they reproduce the correct experimental ordering of the lowest excited states  with
the exception of the missing  $1B_u^+$ state (the HOMO-LUMO excitation), which should lie
\textit{below} the $3A_g^{-}$ in the shorter polyenes such as $\text{C}_8\text{H}_{10}$.  If we perform
a state-averaged DMRG-CASSCF  with 5 states in $\text{C}_8\text{H}_{10}$, we find that
the $1B_u^+$ state lies immediately \textit{above} the $3A_g^-$. This may seem
strange given that CASSCF is generally believed to yield 
qualitatively correct electronic structure, but it reflects the  
wisdom from earlier studies on butadiene that $\sigma$-$\pi$ correlation is very strong
in the $1B_u^+$ state and must be included  to obtain the correct
balance between Rydberg and valence character \cite{cavedavidson,Roos1993,Roos1989,dunningshavitt}. Comparing with the
calculations of Kurashige \etal \cite{kurashige}, which despite having an incomplete
valence active space include dynamic $\sigma$-$\pi$ correlation through
MRMP perturbation theory \cite{hirao_mrmp},  further
indicates that  $\sigma$-$\pi$ correlation  would also lower the
excitation energies of our other excited states.

To better understand the effect of using a complete $\pi$ valence
space on the excitation energies, we have performed some small
benchmark CASSCF calculations on $\text{C}_{12}\text{H}_{14}$ with
$4-12$  active orbitals. These results are presented in
Fig. \ref{fig:activespace}. 
As can be seen, there is a very strong dependence of the excitation energies on the size
of the active space, and even the order of the excitations
changes. Thus, while  an incomplete valence active space  can yield
an excited state ordering in better agreement with experiment, one is tempted
to argue that it does not do so for the right reason.

% states
% changes. While it is possible to pick a small active space which
% yields an excited state ordering in agreement with experiment, it is
% clear from the strong active space size dependence that one should use
% at least the full valence space to obtain the right answer for the
% right reason.

% From the variation of the CASSCF and the CASCI excitation energies we can see the effect of the
% relaxation energy for each excitation. The effect is least predominant in the ground state and most
% in the first excited $2^1A_g^-$ state. Thus the orbital rotation greatly improves the first excitation
% energy $1A_g^- \rightarrow 2A_g^-$. However, the $1A_g^- \rightarrow 1B_u^-$ is still greatly
% over-estimated.

% To see the effect of the active space, calculations with different active spaces were done on
% $C_{12}H_{14}$. Since, the $A_g^-$ states have the maximum multi-reference character, it is the
% $A_g^-$ excited states that are most affected by the change in the active space. With the
% inclusion of the complete active space, the activation energies become comparable to experiment,
% whereas they start from really high unphysical numbers. The ordering of the states are also 
% affected by the increase in active space, since the effect of the active space is not the same on
% all the states.

\begin{figure}[t]
\centering
    \includegraphics[width=3.5in]{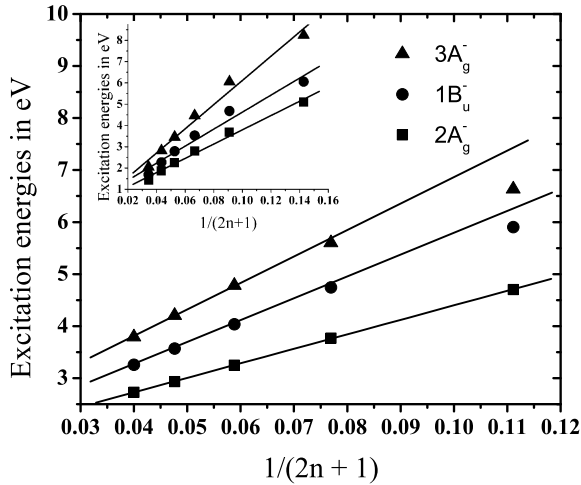}
\caption{\label{fig:linearfit}DMRG-CASSCF excitation energies for the low-lying singlet excited states of 
polyenes ranging from $\text{C}_{12}\text{H}_{14}$ to $\text{C}_{24}\text{H}_{26}$. The excitation energies are plotted against $1/(2n+1)$ where $n$ is the number of double bonds. The ratio of the slopes for the different states is found to be $2:3.0:3.8$ as compared to $2:3.1:3.8$ experimentally. Inset: same plot for the CASCI-MRMP excitation energies from Kurashige \etal \cite{kurashige}. As can be seen, these show a different and less linear-dependence on $1/(2n+1)$.}
\end{figure}

In Fig. \ref{fig:linearfit}, we plot our DMRG-CASSCF excitation energies as a function
of the inverse chain length of the polyenes. Also shown (as an inset)
is the same plot for the excitation energies obtained by 
Kurashige \etal \cite{kurashige}. It is easy to show that in a finite H\"uckel
 model with $n$ sites, the excitation energies have a $\sin(k\pi/2(2n+1))$ chain length
dependence, where $k$ is a quasi-momentum  number that labels the
excitation. For long chains, this implies an asymptotic linear dependence on the
inverse chain length $1/(2n+1)$. Tavan and Schulten conjectured that
this asymptotic behaviour held also in \textit{interacting} systems, and presented evidence from MRD-CI calculations on short-chain Hubbard
 ($n$ up to 7)  and Pariser-Parr-Pople
models ($n$ up to 8) to support
the conjecture \cite{tavan2}. The experimental Resonance Raman excitation profiles
from Sashima \etal \cite{sashima} and Furuichi \etal \cite{furuichi} were also
approximately fitted to the same inverse chain length behaviour, although only over
a small range of $n=9-13$. We see from our results that while the
$2A_g^{-}$ and $1B_u^{-}$ excitation energies fit the  asymptotic $1/(2n+1)$
behaviour well, the $3A_g^{-}$ state shows  curvature more
indicative of the sinusoidal dependence expected when $k \sim
2n+1$. This is consistent with interpreting the $3A_g^{-}$ as an
excitation labelled by a larger
quasi-momentum than $2A_g^{-}$. Interestingly, the excitation energies
of Kurashige \etal show quite different chain-length dependence, with
all three states showing much stronger curvature when their excitation
energies are plotted against  $1/(2n+1)$ in Fig. \ref{fig:linearfit} (inlay).
Fitting our excitation energies for 
$\text{C}_{16}\text{H}_{20}$, $\text{C}_{20}\text{H}_{24}$,
$\text{C}_{24}\text{H}_{26}$ ($n=8-12$) to the asymptotic dependence $1/(2n+1)$,
we obtain slopes of 27.67eV, 41.34eV, 52.63eV for the $2A_g^{-}$, $1B_u^{-}$,
$3A_g^{-}$ excitations,  in reasonable agreement with the
experimental slopes of 31.39eV, 49.07eV and 59.63eV.

\begin{table}[h]
 \centering
 \caption{\label{tab:weights} Single particle nature of the polyene
excitations (in \%). For a given excited state
(e.g. $2A_g^-$), the excitation weight of the transition $i\to j$ is
given by $[\langle 1A_g^- | a^\dag_i a_j |2A_g^-\rangle]^2$. The total
excitation weight is the sum of weights for all transitions; $100\%$
indicates that the given excited state corresponds entirely to single
excitations from the ground state.
 The transition labels  $n\rightarrow m^\prime$ are
interpreted as follows: 1, 2, 3 $\ldots$ denote HOMO, HOMO-1, HOMO-2
$\ldots$ natural orbitals, while $1^\prime$, $2^\prime$, $3^\prime$
denote LUMO, LUMO+1,LUMO+2 natural orbitals. As the polyenes increase in length, the total weight of the single excitations in the low-lying states becomes very small, $<16\%$. }
 \begin{tabular}{|c|c|ddddd|}
 \hline \hline State & Excitation & \multicolumn{4}{c}{No. of conjugated double bonds} & \\ 
 &  weight & $4$        & $   6   $    & $   8   $     & $   10  $    & $   12  $  \\
 \hline \hline
$2A_g^-$ & $2 \rightarrow 1^\prime $ & $ 10.9 $ & $8.6$ & $6.6$ & $5.3$ & $4.3$ \\  
         & $1 \rightarrow 2^\prime $ & $  6.7 $ & $5.9$ & $4.8$ & $4.0$ & $3.3$ \\
         & Total                    & $ 20.0 $ & $18.0$ & $15.4$ & $13.5$ & $12.1$ \\ \hline
$1B_u^-$ & $3 \rightarrow 1^\prime $ & $ 14.5 $ & $10.2$ & $7.9$ & $6.3$ & $5.2$ \\
         & $1 \rightarrow 3^\prime $ & $  7.0 $ & $5.6$ & $4.6$ & $3.9$ & $3.3$ \\
         & Total                    & $ 25.3 $ & $21.8$ & $18.6$ & $16.3$ & $14.7$ \\ \hline
$3A_g^-$ & $4 \rightarrow 1^\prime $ & $ 21.3 $ & $12.8$ & $9.3$ & $7.1$ & $5.6$ \\
         & $1 \rightarrow 4^\prime $ & $  8.2 $ & $6.0$ & $4.7$ & $3.8$ & $3.1$ \\
         & Total                    & $ 32.9 $ & $25.0$ & $20.9$ & $18.0$ & $15.9$ \\ \hline
 \end{tabular}
\end{table}

From the one particle transition density matrices we can  analyse
the single-particle character of our excitations. Given the density
matrix element $w_{ij} = \langle \text{g.s.} | a^\dag_i a_j |
\text{excited}\rangle$ where $i, j$ are natural orbitals in the ground state, we define the
weight of the $i\to j$ excitation as $w_{ij}^2$. The total single
excitation weight is then $\sum_{ij} w_{ij}^2$.   In Table
\ref{tab:weights} we give the largest excitation weights and the total
single excitation weights for the low-lying polyene excited states as
a function of the number of conjugated bonds. We see
the $2A_g^-$, $1B_u^-$ and $3A_g^-$ states are dominated by
many-particle excitations from the ground state (i.e. they have small
single-particle excitation weights) 
and indeed  the single-particle character of the
excitations  decreases even more as the chain-length increases. Remarkably,
in $\text{C}_{24}\text{H}_{26}$ only $<16\%$ of  the  excitation character of
these states can be considered to be of a single-particle nature! These results are
consistent with the analysis by Wormer and Dreuw using coupled cluster
and propagator techniques \cite{dreuw2003cfq}.

\subsection{$\beta$-carotene}

 \begin{figure}[h]
 \centering
     \includegraphics[width=9cm]{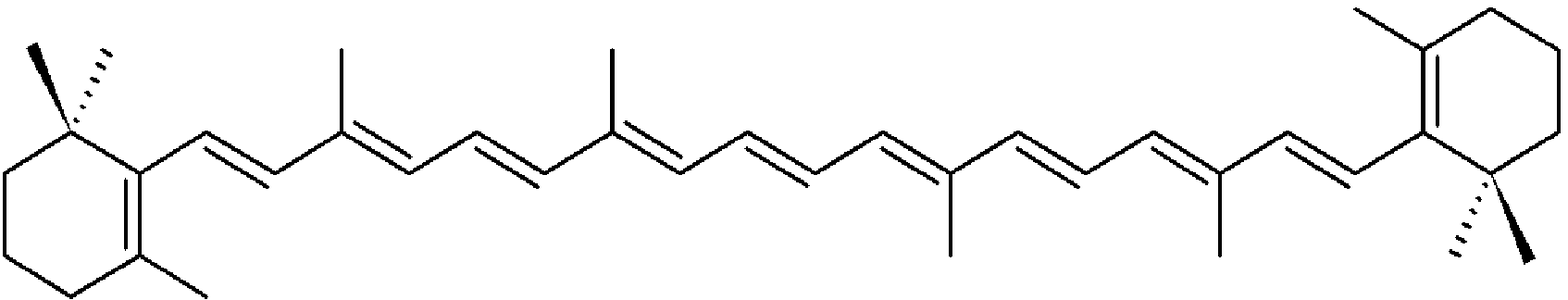}
     \includegraphics[width=9cm]{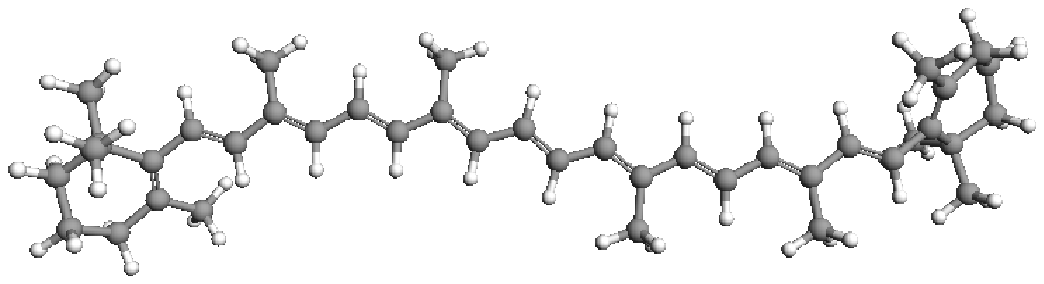}
 \caption{\label{fig:carotene}\textit{s-cis} $\beta$-carotene.}
 \end{figure}

% [assignment of conjugation lengths non-trivial (frank)]
Carotenoids, the family of substituted polyenes,   are the primary light
harvesting pigments in the LH2  complex. 
Light harvesting proceeds by the transfer of  energy from an array of 
 carotenoids  to nearby bacteriochlorophylls and thence to the
photosynthetic centre. Many essential questions remain unanswered as
to the precise mechanism of this energy transfer \cite{sunstrom1,schulten-carotene,fleming,hsuhead-gordon1,hsuhead-gordon, dreuw-carotene,dreuw2003cfq}. 
While the absorption of
light places the carotenoid in the dipole allowed excited state, there
can be a fast  internal conversion to the aforementioned dark states of
the polyene backbone, and thus multiple pathways for energy transfer
to the bacteriochlorophyll.
In carotenoids, the dipole allowed transition is usually labelled
$S2$, while historically the dark state is labelled $S1$. However,
with the discovery, as previously described, of \textit{additional}
dark states below $S2$ in these molecules \cite{cogdellscience,onaka,sashima,sashima2,furuichi,fujii}, this nomenclature can be confusing. An
alternative nomenclature is to simply re-use the polyene excited state
labels, even though the carotenoids have a lower point
group symmetry. We will follow this practice here.

\subsubsection{Discussion}

\begin{table}[h]
  \centering
  \caption{\label{tab:carotene}DMRG-CASSCF energies, symmetries, and oscillator strengths for the lowest lying singlet excited states in $\beta$-carotene with the complete $\pi$-valence (22,22) active space.  Total energies in $E_h$, excitation energies in $eV$, oscillator strengths in $a.u.$. The estimated error of the
  DMRG-CASSCF energies from the exact CASSCF energies in the same
  active space is less than 0.1m$E_h$. Oscillator strengths are  for the ground-state, excited state transition. }
  \begin{tabular}{|c|c|c|c|c|}
  \hline\hline  
Symmetry & DMRG-CASSCF     & Excitation & Oscillator           & Expt\\
         & total energy     &  energy    &  Strength      &  \\
  \hline \hline
$1A_g^-$  & $-1546.914545$ &        &            & \\
$2A_g^-$  & $-1546.804503$ & $2.99$ &  Forbidden & $1.81$ \footnotemark[1]\\
$1B_u^-$  & $-1546.781125$ & $3.63$ &  $0.2025$ & $2.05$ \footnotemark[1] \\
$3A_g^-$  & $-1546.755822$ & $4.31$ &  Forbidden & $(2.22)$ \footnotemark[2] \\
\hline\hline
  \end{tabular}
\footnotetext[1]{\cite{sashima2}.} 
\footnotetext[2]{Excitation measured for lycopene \cite{furuichi}.} 
\end{table}

\begin{figure}
  \begin{center}
    \includegraphics[width=3.5in]{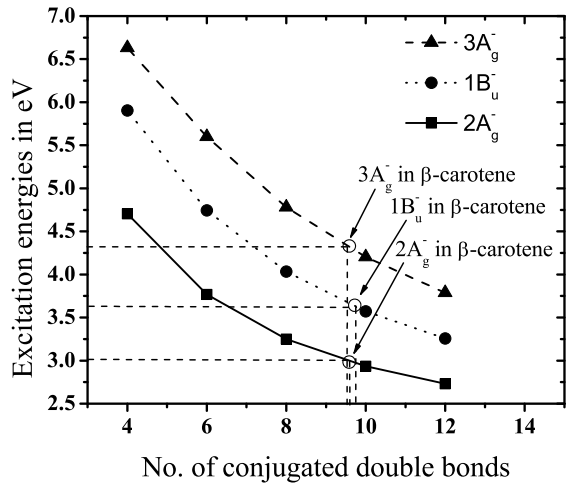}
    \caption{\label{fig:doublebond} Polyene and carotene excitation energies vs the number of double bonds: the $\beta$-carotene
excitation energies when fitted to the polyene excitation energies give an effective conjugation length of $9.5-9.7$.}
  \end{center}
\end{figure}

\begin{figure}[h]
\centering
\subfigure[LUMO+1 natural orbital]
{
    \includegraphics[width=3.3in]{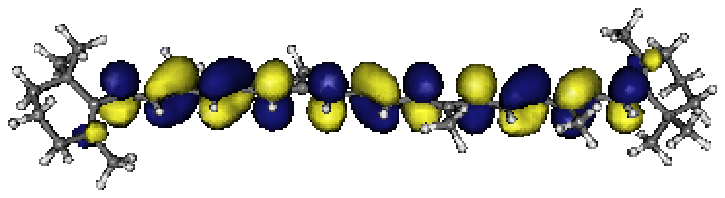}
}
\\
\subfigure[LUMO natural orbital]
{
    \includegraphics[width=3.3in]{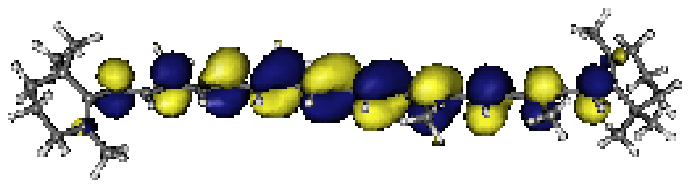}
}
\subfigure[HOMO natural orbital]
{
    \includegraphics[width=3.3in]{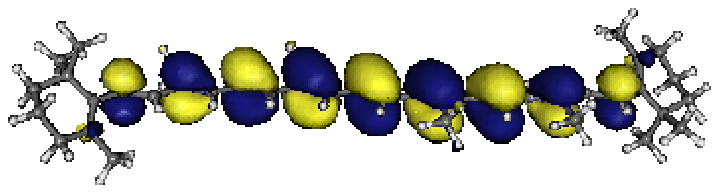}
}
\subfigure[HOMO-1 natural orbital]
{
    \includegraphics[width=3.3in]{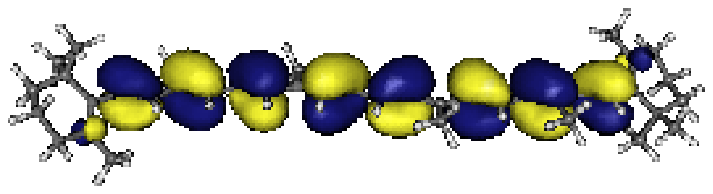}
}
\caption{\label{fig:nos} Natural orbitals corresponding to the HOMO-1 through LUMO+1 states.\\
These orbitals participate in the lowest lying singlet excitations in $\beta$-carotene and contain little density on the non-planar end groups.}
\end{figure}

%% [Results are similar to the polyenes ordering and missing state and
%% overestimate. Results are qualitatively similar to expt]

We have chosen  to study \textit{s-cis} $\beta$-carotene (see
Fig. \ref{fig:carotene}) as a representative carotenoid. 
It is the dominant natural conformer although the \textit{all-trans} form is also studied. 
Crystalline $\beta$-carotene  has
$C_i$ symmetry with a conjugated backbone that lies almost entirely on the
$xy$ plane except for end groups which are twisted out of
plane \cite{Schlucker772,Berezin771}. (In the biological setting, carotenoid pigments usually adopt a
twisted configuration in the conjugated backbone\cite{Wang770,Qian757}). There are 11
conjugated double bonds in the backbone.  Our study employed the same
calculation procedure as described in Sec. \ref{sec:polyene_compute}
with the exception that we used a 6-31G basis set in the DMRG-CASSCF calculation due to
the large size of the molecule. State-averaged DMRG-CASSCF
calculations were performed with 4 states and a (22,22) complete
$\pi$-valence space, in the manner described in Sec. \ref{sec:polyene_compute}.

In Table \ref{tab:carotene} we present the energies, symmetries, and oscillator
strengths for the ground state
and first 3 excitations in $\beta$-carotene. We reproduce the
 state ordering $1A_g^- < 2A_g^- < 1B_u^- < 3A_g^-$
as assigned by Furuichi \etal \cite{furuichi} (note that the $1B_u^+$ which does not
appear in our calculation  indeed lies above the
$3A_g^-$ state in this molecule). However, just as in the polyenes,
the excitation energies from the DMRG-CASSCF procedure are generally
overestimated in comparison with experiment, most likely due to the lack of
$\sigma$-$\pi$ dynamic correlation. 

% Results are similar to the polyenes and we do get the correct ordering \cite{Furuichi} of excited states.
% The orbital optimization has the maximum effect on the $1B_u^-$ state as is expected and the least on the
% ground state. Fluorescence spectroscopic study by Onaka et al \cite{onaka} shows that
% the $S_0 \rightarrow S_1$ and $S_0 \rightarrow S_2$ transition is 1.79eV and 2.53 eV in n-hexane. Our
% results show a qualitative agreement to this result. Unlike the TDDFT studies which give lower excitation
% energies than the experimental value and CIS which give different ordering ($S_0$ lower than $S_1$),
% CASSCF gives the correct ordering of the states and over-estimated excitation energies.

%% \begin{figure}[t]
%% \centering
%%     \includegraphics[width=10cm]{pic/carotene_no11.eps}
%% \caption{\label{fig:no2} Highestest Occupied Natural Orbital}
%% \end{figure}

%% \begin{figure}[t]
%% \centering
%%     \includegraphics[width=10cm]{pic/carotene_no12.eps}
%% \caption{\label{fig:no1} Lowest Unoccupied Natural Orbital}
%% \end{figure}

%% \begin{figure}[t]
%% \centering
%%     \includegraphics[width=10cm]{pic/carotene_no13.eps}
%% \caption{\label{fig:no2} Lowest Unoccupied Natural Orbital + 1}
%% \end{figure}

%% [Effective conjugation length by comparing with the polyenes. and ref
%% to Onaka where they get 9.7 ... The twisted end groups etc. Which is
%% also shown by the electron density HONO, LUNO.]

A question that has received some attention in the literature is the
effective conjugation length of carotenoids, since the presence of
substituents and non-planar geometries are expected to modify this
from the naive value deduced from the Lewis structure
\cite{frankscites}. Formally,
$\beta$-carotene has 11 double bonds in the polyene backbone, but by comparing the
excitation energies of the  polyenes with our $\beta$-carotene
excitation energies, we can estimate a reduced conjugation length of
9.5-9.7 bonds, which is
very close to the experimental estimate of 9.7 of Onaka \etal \cite{onaka}. This
reduced conjugation length  results from the twist in the carotene end-groups. In
Fig. \ref{fig:nos} we plot the DMRG-CASSCF natural orbitals corresponding to the HOMO, HOMO-1, LUMO,
and LUMO+1. As can be seen, there is very little density in these
orbitals  on the carotene end-groups, and this is consistent with our
reduced effective conjugation length.

\section{\label{sec:level5}Conclusion}

% In quantum chemistry, the traditional methods for difficult
% multireference problems is the Complete Active Space Self-Consistent
% Field (CASSCF) method, where the electronic structure is solved
% exactly within a small space of optimised active orbitals. 

In this work, we described how to efficiently implement orbital
optimisation using the Density Matrix Renormalization
Group (DMRG) wavefunction. We have named the resulting method
DMRG-CASSCF, and by virtue of the compact nature of the DMRG
wavefunction, this now enables us to handle much larger active
spaces than are possible with the traditional CASSCF algorithm. 
As a sample application, we have used our DMRG-CASSCF implementation
to study the low-lying excitations of polyenes from
$\text{C}_8\text{H}_{10}$ to $\text{C}_{24}\text{H}_{26}$ as well as
the light-harvesting pigment $\beta$-carotene, with up to a (24,24) complete active space.
Our calculations reproduce the state ordering of the dark states that
have been recently observed by  Resonance Raman studies. However, as
expected from earlier CASSCF studies, the energy of the optically allowed HOMO-LUMO
 $1B_u^+$ transition is still overestimated, as a result of the  lack of dynamic $\sigma$-$\pi$ correlation in the
DMRG-CASSCF method. We therefore view  the  incorporation of dynamic correlation, either via perturbation theory
or via canonical transformation \cite{whitecd, yanaict1} into the DMRG-CASSCF method to
present an important next direction for development.

% While we have previously shown how the DMRG
% enables  the near-exact treatment of large active spaces in a number of
% difficult problems, this required , it has not previously been possible to optimise
% the active orbitals, thereby limiting such calculations to problems
% where the active degrees of freedom are readily identified. The
% orbital optimised DMRG   calculationspreceding works have shown how
% the Density Matrix Renormalization Group The resulting method, which we call
% DMRG-CASSCF,  allows us to treat difficult multireference problems
% with large active spaces, even when the active orbitals are not easily
% identified. 

% This allows us to now use the DMRG to
% perform Complete-Active-Sp

% While the Density Matrix Renormalization Group has enabled the
% accurate description of multireference problems in large active
% spaces, 

% Using our DMRG-CASSCF method we could study the low-lying excitations
% in 

% we found

% Something else -- canonical transformation
\section{Acknowledgments}

This work was supported by 
Cornell University, the Cornell Center
for Materials Research (CCMR), the David and
Lucile Packard Foundation, the National Science
Foundation CAREER program CHE-0645380, the Alfred P. Sloan
Foundation, and the Department of Energy, Office of Science through
award DE-FG02-07ER46432.
Johannes Hachmann would like to acknowledge support provided by a Kekul\'{e} Fellowship of
the Fond der
Chemischen Industrie.

\bibliographystyle{aip}
\bibliography{dmrgscf_theory}
\end{document}